\documentclass[aps,pra,twocolumn,superscriptaddress,floatfix]{revtex4-1}

\usepackage{bm}
\usepackage[colorlinks=true, citecolor=blue, urlcolor=blue]{hyperref}\usepackage{epsf,graphics,graphicx}
\usepackage{amsmath}
\usepackage{latexsym}
\usepackage{amssymb}
\usepackage{graphicx}
\usepackage{color}
\usepackage{soul,xcolor}
\setstcolor{red}
\usepackage{ulem}
\usepackage{cancel}
\usepackage{tikz}
\usepackage[export]{adjustbox}
\usepackage{subcaption}
\usepackage{braket}
\usepackage{multirow}
\usepackage{caption}
\usepackage{float}
\usepackage{lipsum}

\usepackage{epsfig}
\input{epsf}

\begin{document}
\title{Fano resonances in tilted Weyl semimetals in an oscillating quantum well}
\author{Souvik Das$^\clubsuit$}
\affiliation{Indian Institute of Science Education and Research Kolkata, India} 
\author{Arnab Maity$^\clubsuit$}
\affiliation{Indian Institute of Science Education and Research Kolkata, India} 
\author{Rajib Sarkar$^\clubsuit$}
\affiliation{Indian Institute of Science Education and Research Kolkata, India} 
\author{Anirudha Menon}
\affiliation{School of Physical Sciences, Indian Association for the Cultivation of Sciences, Kolkata, India} 
\author{Tanay Nag}
\affiliation{Department of Physics, BITS Pilani-Hyderabad Campus, Telangana 500078, India}
\email{tanay.nag@hyderabad.bits-pilani.ac.in}
\affiliation{Department of Physics and Astronomy, Uppsala University, Box 516, 75120 Uppsala, Sweden}
\author{Banasri Basu}
\affiliation{Physics and  Mathematics Unit \& Interdisciplinary Statistical Research Unit, Indian Statistical  Institute, Kolkata, India} 
\thanks{$^\clubsuit$SD, AM and RS contributed equally.}

\date{\today}

\begin{abstract}
	 Considering the low-energy model of tilted Weyl semimetal, we study the electronic transmission through a periodically driven quantum well, oriented in the transverse direction with respect to the tilt. We adopt  the  formalism of Floquet scattering theory and  investigate the emergence of Fano resonances as an outcome of matching between the Floquet sidebands and quasi-bound states. The Fano resonance energy changes linearly with the tilt strength suggesting the fact that tilt-mediated part of quasi-bound states energies depends on the above factor. Given a value of momentum   parallel (perpendicular) to the tilt, we find that the energy gap between two Fano resonances, appearing for two adjacent values of transverse (collinear)  momentum with respect to the tilt direction, is insensitive (sensitive) to the change in the tilt strength. Such a coupled (decoupled) behavior of tilt strength and the collinear (transverse) momentum can be understood from the tilt-mediated and normal parts of the quasi-bound state energies inside the potential well. We vary the other tilt parameters and chirality of the Weyl points to conclusively verify the exact form of the tilt-mediated part of the quasi-bound state energy that is the same as the tilt term in the static dispersion. \textcolor{black}{The tilt orientation can significantly alter the transport in terms of evolution of Fano resoance energy with tilt momentum.}
	 We analytically find the explicit form of the bound state energy that further supports all our numerical findings. Our work paves the way to probe the tilt-mediated part of quasi-bound state energy to understand the complex interplay between the tilt and Fano resonance.
\end{abstract}

\maketitle

\begin{figure*}[ht]
\includegraphics[width=0.99\textwidth]{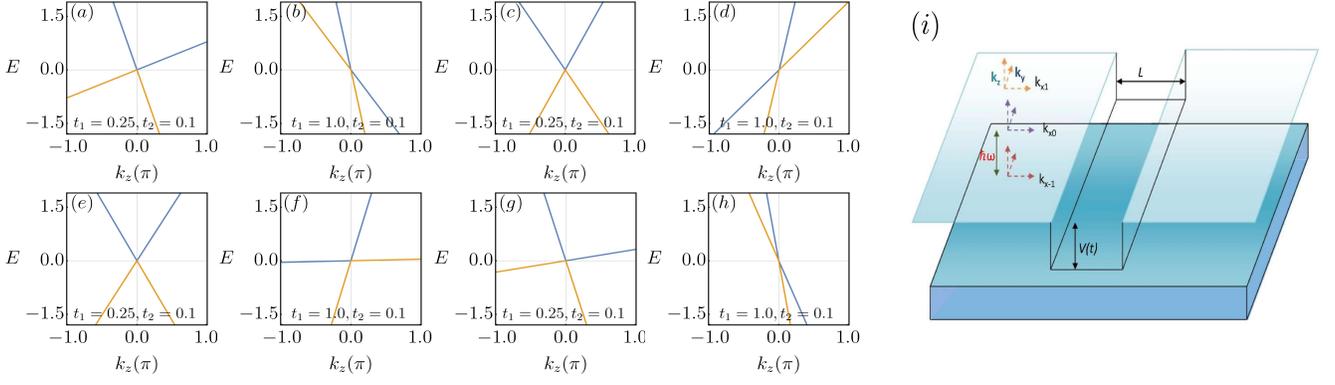}
\caption{
{The dispersion for the Hamiltonian in Eq. (\ref{00}) are shown in (a-h). We consider $(\phi_1,\phi_2)=(\pi/4,\pi/2)$ and $(\pi,\pi/2)$  in (a,b) and (c,d) [(e,f) and (g,h)], respectively for $s=+1$ [$s=-1$]. The yellow and blue bands represent $E_{s,-}$  and $E_{s,+}$ as given in Eq. (\ref{eq:energy}). We choose $t=t_z=1$, $k_x=k_y=0$ and $k_0=\pi/2$.  (i):
The potential landscape is shown  where the electron is transported in the $x$-direction through a harmonically driven potential $V(t) = -V_0 + V_1\cos(\omega t)$ of width $L$. This potential landscape is only applicable along $x$-direction which is perpendicular to the direction of tilt. $\hbar \omega$ represents the photon energy for virtual processes, and $k_{xi}$ represent the various momenta for the Floquet sidebands with $i=-1,0,1$.}}
\label{1}
\end{figure*}

\section{Introduction}
\label{sec1}

Weyl semimetals (WSMs) have been the topic of extended research in condensed matter physics over the past decade \cite{new3,new5,new6,new7,arp1,arp2}, with the proposal for and discovery of the Weyl fermion. 
This led to a plethora of theoretical as well as experimental studies focussing on the  features, characteristics, and transport properties of such materials \cite{Yan17}. The minimal model of a time reversal symmetry broken WSM  consists of two Weyl points which act as a source and sink for Berry curvature in momentum space. These points are usually observed near the Fermi surface where the lightcone-like dispersion of the WSM meets \cite{new3,new4,new5,new6,new7,11,12}.  The first proposed WSMs are now defined in the literature as the type-I WSM, which have a conical dispersion in momentum space. The addition of a Lorentz symmetry violating tilt term to the WSM Hamiltonian gives rise to a new type of WSM called the type-II WSM \cite{WSMII,FP1,FP2}. The type-II phase arises when the value of tilt generally exceeds the Fermi velocity, and this phase has some features which are distinct from the type-I phase including an extended Fermi surface.
WSMs, irrespective of their type, typically exhibit a large variety of exotic properties  
like negative magneto-resistance \cite{NMR}, universal quantum Hall signatures \cite{new5,AM1,Nag21,Sadhukhan23,Xiong22}, and the chiral magnetic effect \cite{Nag21}, all of which are some way or the other
consequences of the chiral anomaly. Importantly, 
the transport properties of type-II WSMs are starkly different from those of the type-I phase \cite{Sadhukhan21,Sadhukhan21a,Das21,Nag22}. This can be ascribed to marked differences in the density of states at the Fermi level \cite{WSMII}. To add more, type-II WSMs possess novel quantum oscillations due to momentum space Klein tunneling \cite{T21}, and a modified anomalous Hall conductivity \cite{T22}. Tilting of the Weyl cones also affects the Fano factor as described in \cite{T23}. Therefore, one can infer that the tilt-mediated transport phenomena have attracted a lot of attention in recent years.

The study of periodically driven systems has always been of great interest to physicists because of the potential to alter material behavior \cite{oka09, kitagawa10, rudner13, nathan15, titum16, po16, kar18,Nag14,Nag19,Tamang21,Nag21b,Kundu21} and obtain phenomena which may be controlled in a laboratory environment. Periodic driving is best described by employing the Floquet formalism \cite{F1,F2,F3}, which is the temporal analog of Bloch's theorem in frequency space. The technique utilizes the time periodicity of the driving potential to express the wavefunction as the product of a phase and a periodic function in time, reducing the complexity  to solve  the time-dependent Schr\"odinger equation. Advances in Floquet methods have led to the development of Floquet perturbation theory in the form of the high-frequency and van-Vleck expansions \cite{AM2,AM3} where an effective time-independent Floquet Hamiltonian can be obtained. Such methods have been applied to Dirac semimetals \cite{FD1}, WSMs \cite{AM2,AM3}, and even quantum spin liquids \cite{FQSL}, to generate controllable transport properties. 

In the context of transmission spectra,
Fano resonances have been studied in many areas of physics since its inception in 1935 \cite{Z2}. It is characterized by a perfect transmission followed by a total reflection or vice-versa, leading to an asymmetric resonance profile. The phenomenon can be explained by considering the interaction of localized states with a range of propagating continuum modes, wherein the transmission enhancement occurs due to constructive interference, and the reflection enhancement due to destructive interference, between different quantum trajectories. In the context of non-adiabatically driven quantum systems, Floquet sidebands are formed, and when one of these approaches a quasi-bound state, a Fano resonance occurs. Fano resonances have been noticed in various contexts such as for light propagation in photonic devices \cite{Z1,Z3}, scattering in mesoscopic transport systems \cite{Z1,Z4,Z5,Z6,Z7,Z8,Z9,Z10,Z11,Z12,Z13,Z14} and superconducting Josephson junctions \cite{Z1,Z15}, graphene \cite{Z16}, Dirac semimetals \cite{Z17}, and WSMs \cite{FW1}.

Previous literature has dealt with the Fano resonance structure of Dirac \cite{Z17} and Weyl systems \cite{SB1} employing Floquet theory in the presence of a harmonically driven potential well. However, there is not much commentary on the consequences of tilt in the energy dispersion 
on Fano resonances in these cases \cite{Gregefalk23}. Considering the fact that tilt appears to be a key factor in manipulating the transport properties of WSMs as described previously \cite{T21,T22,T23}, and motivated by the intricacies of Fano resonances, we examine the role of tilting on Fano resonances in three-dimensional WSMs in this work. In particular, we examine the Fano resonances, exploiting the Floquet scattering formalism, as a function of tilt strength on type-I and type-II WSMs and uncover hitherto unexplored physics in this context where tilt direction (say along $z$-axis) is transverse to the transport through the harmonically driven potential well (say along $x$-axis). Given a value of tilt momentum,
we find that the Fano resonance energies depend linearly on tilt strength while  the energy gap between two Fano resonances, appearing for two adjacent values of transverse momentum (say along $y$-axis)
remains insensitive to the change in the tilt strength.
On the other hand, given a value of transverse momentum, Fano resonance energies depend  linearly on 
tilt strength and the corresponding tilt momentum in a coupled manner
such that the energy gap between two  Fano resonances, occurring for two adjacent values of tilt momentum, changes  with the variation in the tilt strength.  The seclusive (inclusive) behaviour between tilt strength and the transverse (collinear) momentum with respect to tilt direction is essentially caused by tilt-mediated  and normal parts of the quasi-bound state energies inside the potential well. We also investigate the effect of chirality on the transmission spectra in the above cases. This allows us to understand the interplay between the tilt and 
Floquet scattering leading to 
the rich profile of Fano resonances through a harmonically driven potential well for 3D WSMs.

The structure of this manuscript is as follows: In Sec. \ref{sec2} we consider the minimal model of a time reversal symmetry broken WSM with tilt, capable of hosting the type-I and type-II phases, and introduce the harmonically driven quantum well. Sec. \ref{sec3} contains the solution to Schrodinger's equation and formulation of the S-matrix. We present the results for the transmission coefficients in Sec. \ref{sec4} for varying tilts, where the characteristic Fano resonances are demonstrated. A discussion on the results and the implications on future research is presented in Sec. \ref{sec5}. We summarize our findings and conclude in Sec. \ref{sec6}.

\section{Model and time-dependent quantum well}
\label{sec2}

We begin our study by considering the low-energy minimal model Hamiltonian around a  Weyl point of chirality $s$ as given below  \cite{Nag22}  
\begin{align}\label{00}
	H=2k_z[t_1\sin(\phi_1-sk_0 )+2t_2 \sin(\phi_2 -2sk_0)]\sigma_0 \nonumber \\
	+t(\sigma_xk_x+\sigma_yk_y)+st_z\sigma_zk_z\sin k_0,
\end{align}
where $\sigma_{x}=\begin{pmatrix}
	0 & 1 \\
	1 & 0 \\
\end{pmatrix}$,  $\sigma_{y}=\begin{pmatrix}
	0 & -i\\
	i & 0 \\
\end{pmatrix}$ and $\sigma_{z}=\begin{pmatrix}
	1 & 0 \\
	0 & -1 \\
\end{pmatrix}$ are Pauli matrices. $t_z$ and $t$ are the spatially non-isotropic Fermi velocities of the Weyl points. Here the identity term causes a tilt along the $k_z$ direction where $t_{1,2}$ denotes the parameters which control tilt strength. The phase factors $\phi_{1,2}$ represent the fluxes in hopping along $z$-direction in a lattice regularized model where the Weyl points appear at $(0,0,s k_0)$. Note that the tilt term can be finite even when $\phi_{1,2}=0$ indicating the fact that finite $t_{1,2}$  is essentially  responsible for a tilted dispersion.  \textcolor{black}{The above model Eq. (\ref{00}) can host type-I, type-II as well as hybrid phase where one Weyl point is type-I and  its chiral counterpart is type-II. Once both the Weyl points can either be type-I or type-II, only one set of parameters, comprised of ($t_1,\phi_1$) or ($t_2,\phi_2$) only, is enough to give rise to the above phases where both the Weyl points are tilted in an identical manner usually. Importantly, for the hybrid phase, one usually needs two sets of parameters ($t_1,\phi_1$) and ($t_2,\phi_2$) such that two Weyl points can be tilted in different manner.  However, the hybrid phase can only be supported by the lattice version of the model. In the present case, we restrict ourselves to the type-I and type-II phases as we consider an isolated Weyl point individually.  Therefore, once can in principle adopt a simple form of the tilt term with only one set of parameters. We will demonstrate the upshot of the tilt term in the Fano resonance energy in Fig.~\ref{fig:6} that may not be obvious from the simple form of tilt term.}

The degree of tilt $\eta_s$ for a given Weyl point of chirality $s$ is defined as 
\begin{equation}\label{e3}
	\eta_s=\Bigg{|}\frac{2[t_1\sin(\phi_1-sk_0 )+2t_2 \sin(\phi_2 -2sk_0)]}{st_z \sin k_0}\Bigg{|}.
\end{equation}
This can be evaluated separately for each node to get the type-I ($\eta_s < 1$) and type-II ($\eta_s > 1$) WSM phases in general with $s=\pm 1$. 
$\eta_s = 1$ marks a Lifshitz phase transition which alters the nature of the Fermi surface as has been observed in the two phases.
The dispersion relation, associated with the Hamiltonian Eq. (\ref{00}), is given by

\begin{align}
	E_{s,u} =& 2k_z[t_1\sin(\phi_1-sk_0 )+2t_2 \sin(\phi_2 -2sk_0)] \nonumber \\ &+u\sqrt{t^2(k_x^2 + k_y^2) + t_z^2k_z^2 \sin^2 k_0},
	\label{eq:energy}
\end{align} 
where $u=\pm$ represent the conduction and valence bands. The  Hamiltonian hosts  both type-I and type-II phases of WSMs as depicted in Figs. \ref{1} (a-h). The band dispersion $E_{s,u}$ along the tilt momentum $k_z$ for type-I and type-II phases are depicted in Figs. \ref{1} (a,c,e,g) and (b,d,f,h), respectively.

We now choose to subject this system to a harmonically driven quantum well of width $L$ in the $x$-direction. The potential $V(t)$ for the well is defined as \cite{Fano1,Z16,Gregefalk23}, 
\begin{equation}
	V(t)= \begin{array}{cc}
		\Bigg{\{} & 
		\begin{array}{cc}
			-V_{0}+V_{1}\cos(\omega t ) & |x| \leq L/2  \\
			0 & |x| > L/2
		\end{array}
	\end{array}
\end{equation}

where $V_0$ is the depth of the static potential well. $V_1$ and $\omega$ denote the amplitude and frequency of the harmonic drive.    
The setup of our system is shown in Fig. \ref{1} (i), where the  qualitative features of the driving potential in the three regions of interest are shown explicitly. The harmonically driven quantum well is confined to a region of width $L$ in the $x$-direction, while such a potential landscape is absent  
in the $z,y$-directions. Note that there is no potential outside of the specified region $-L/2 <x <L/2$.  The electron wavefunction is simply plane waves along the $z,y$-directions while $x$-component is non-trivial. It is very important to note that the tilt and the direction of flow of electrons are mutually perpendicular as this will have a significant role to play in our forthcoming results.

\section{Floquet S-matrix}
\label{sec3}


Let us define the wave function as a two-component spinor $\psi=\begin{pmatrix}\psi_1 \\ \psi_2 \end{pmatrix}$, and with this, we now  solve Schr\"odinger's equation in the three regions described above employing the Floquet ansatz. First we consider the region $|x| > L/2$ and find an analytical expression for the wavefunction as follows
\begin{equation}
	\psi=\begin{pmatrix}
		1\\ S_\pm(E)
	\end{pmatrix}\exp{[\pm ik_x x + ik_y y+ ik_z z]},
\end{equation}\par
where \textcolor{black}{$S_\pm (E)=\frac{t(\pm k_x+ik_y)}{E-\tilde{t} k_z+st_z k_z\sin k_0}$ with $\tilde{t}=2t_1\sin(\phi_1-sk_0 )+4t_2 \sin(\phi_2 -2sk_0)$} and $E$ represents the incident energy.
\textcolor{black}{This term is regarded as normalization of the wave-function. It is hence implied that $S_{\pm}(E)$ is a function of incident energy $E$,  momentum  $k_{x,y,z}$, tilt factor $\tilde{t}$, and $t_z$. However, for the ease of notation, we adopt the notation $S_\pm (E)$.}
In order to obtain the wavefunction for the potential region $-L/2\leq x \leq L/2$, we replace $E$ with $E-V_0$.
This results in the substitution of  
$k_x$ and  $S_{\pm}(E)$ respectively with $q_x$ and $S^{'}_{\pm}(E)=S_{\pm}(E-V_0)$ within the above region. Solving the Schr\"odinger equation we find that the wavefunction for all spatial regions is \cite{Fano1,Z16}

\begin{widetext}
\begin{equation}
	\psi_n=e^{-iEt/\hbar +ik_yy+ik_zz} \\
	\begin{cases}
		\begin{aligned}
			&A^i_nN_{n_+}\begin{pmatrix} 1\\ S_+(E_n)\end{pmatrix} e^{ik_{xn} x} +A^o_nN_{n_-}\begin{pmatrix}1\\ S_-(E_n)\end{pmatrix}e^{-ik_{xn}x} ,&  x<-L/2 \\
			&\sum_{m=-\infty}^{\infty}
			[a_mN'_{m_+}\begin{pmatrix} 1\\ S'_+(E_m) \end{pmatrix}e^{iq_{xm} x}+b_mN'_{m_-}\begin{pmatrix} 1\\ S'_-(E_m)\end{pmatrix}e^{-iq_{xm} x}]J_{n-m} \bigg{(} \frac{V_1}{\hbar\omega} \bigg{)}  ,& \hfill -L/2\leq x\leq L/2\\&
			B^i_nN_{n_-}\begin{pmatrix} 1\\ S_-(E_n) \end{pmatrix}e^{-ik_{xn}x} +B^o_nN_{n_+}\begin{pmatrix} 1\\S_+(E_n)\end{pmatrix}e^{ik_{xn}x} ,& x>L/2\end{aligned}\end{cases}
   \label{eq_wfall}
\end{equation} 
\end{widetext}
Here $N_{n_\pm}=\frac{1}{\sqrt{1+|S_{\pm}(E_n)|^2}}$ and $N'_{m_\pm}=\frac{1}{\sqrt{1+|S'_{\pm}(E_m)|^2}}$ are the appropriate normalizing factors,  and $A_n^{i(o)}$ and $B_n^{i(o)}$ are the amplitudes of the incoming (outgoing) waves, associated with $n$-th channel, at the left and right boundaries of the well.
\textcolor{black}{Note that $E_n=E + n\omega$ considering $\hbar=1$. In Eq. (\ref{eq_wfall}), one can thus find $S_\pm (E_n)=\frac{t(\pm k_x+ik_y)}{E_n-\tilde{t} k_z+st_z k_z\sin k_0}$, $S'_\pm (E_n)=\frac{t(\pm k_x+ik_y)}{E_n-V_0-\tilde{t} k_z+st_z k_z\sin k_0}$.}


From fundamental quantum mechanical considerations, we demand continuity and differentiability of the wave function at the boundaries of the driving potential. This allows us to relate the 
incoming amplitudes with the outgoing amplitudes via the intermediate $a_n$ and $b_n$ amplitudes. The detailed derivation on the application of the boundary conditions is given in Appendix \ref{app1}. We proceed with the construction of the scattering matrix $S$. The scattering matrix is defined as \cite{Blanter2000}, 
\begin{equation}
\begin{pmatrix}
	A^o\\
	B^o
\end{pmatrix}=\begin{pmatrix}
	R & T^{'}\\
	T & R^{'}
\end{pmatrix}\begin{pmatrix}
	A^i\\
	B^i
\end{pmatrix}
=S\begin{pmatrix}
	A^i\\
	B^i
\end{pmatrix}
\end{equation}
where $A^i$ and $B^i$ represent the amplitudes of the incident waves and $A^o$ and $B^o$ represent the amplitudes of the outgoing signal. The reflection ($R,R^{'}$) and transmission ($T,T^{'}$) sectors of the $S$ matrix can be represented as \cite{Fano1}
\begin{align}
\begin{pmatrix}
	R & T^{'}\\
	T & R^{'}
\end{pmatrix}
=
\begin{pmatrix}
	r_{00} & r_{01} & .. & t^{'}_{00} & t^{'}_{01} & ..\\
	r_{10} & r_{11} & .. & t^{'}_{10} & t^{'}_{11} & ..\\
	.      & .      & .. & .          & .          & .. \\    
	.      & .      & .. & .          & .          & .. \\
	t_{10} & t_{11} & .. & r^{'}_{10} & r^{'}_{11} & ..\\
	t_{10} & t_{11} & .. & r^{'}_{10} & r^{'}_{11} & ..\\
	.      & .      & .. & .          & .          & .. \\    
	.      & .      & .. & .          & .          & .. \\,
\end{pmatrix}
\end{align}
where $r_{nm}$ and $t_{nm}$ are reflection and transmission amplitudes for modes incident from the left and $r^{'}_{nm}$ and $t^{'}_{nm}$ are reflection and transmission amplitudes for modes incident from the right. The detailed derivation on S-matrix  formulation is given in Appendix \ref{app2}.
Here, we have $n,m \in (0,\infty)$ since only propagating modes are considered. Note that we consider only one incoming electron of energy $E$ i.e., there exists only one incoming channel. Hence, we only work with $n=0$ case. From the $S$ matrix, we can define the total transmission coefficient $T$ as \cite{Z16, Gregefalk23,Fano1}

\begin{align}
	T = \sum_{m=0}^{\infty}\frac{{\rm Re}(k_0)}{{\rm Re}(k_m)}|t_{0m}|^2.
	\label{eq:transmission}
\end{align}
\textcolor{black}{ $m$ in Eqs. (\ref{eq_wfall}) and (\ref{eq:transmission}) are the same index on which the summation has to be performed.  Note that $m$ ($n$) represents Floquet sidebands  (incoming channels) within (outside) the potential region
in Eq. (\ref{eq_wfall}) whereas in Eq. (\ref{eq:transmission}), $m$ represents the column index of the transmission matrix $t_{nm}$. }

\section{Results}
\label{sec4}

\begin{figure*}[t]
		
		\centering
		\includegraphics[scale = .55]{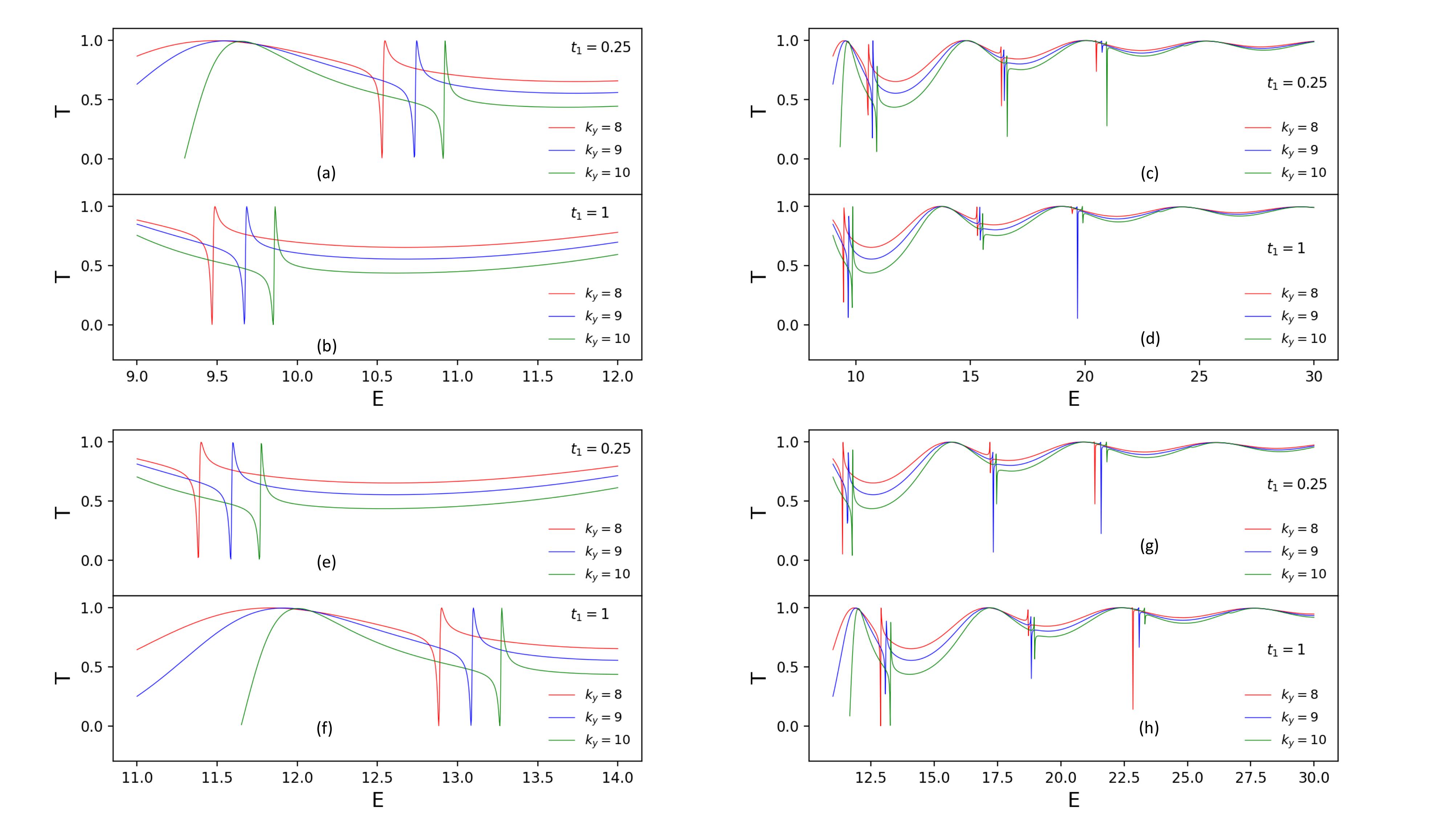}
			
	\caption{Transmission spectra $T$, computed from Eq. (\ref{eq:transmission}), as a function of incident energy $E$ for positive chirality Weyl point with $s=+1$ are shown above keeping $k_z$ fixed while changing value of $k_y$. Here (a,b) depict first Fano resonance and (c,d) show a large region of incident energy region for $\phi_1=\pi/4$ and $\phi_2=\pi/2$. We repeat the plots (a,b) and (c,d) in (e,f) and (g,h), respectively for $\phi_1=\pi$ and $\phi_2=\pi/2$. Parameters are taken as follows $L=0.6$, $k_z=1$, $V_0=100$, $V_1=1$ and $\omega=15$.}
	\label{fig:1}
		
\end{figure*}

\begin{figure*}[t]
		
			\centering
			\includegraphics[scale = .55]{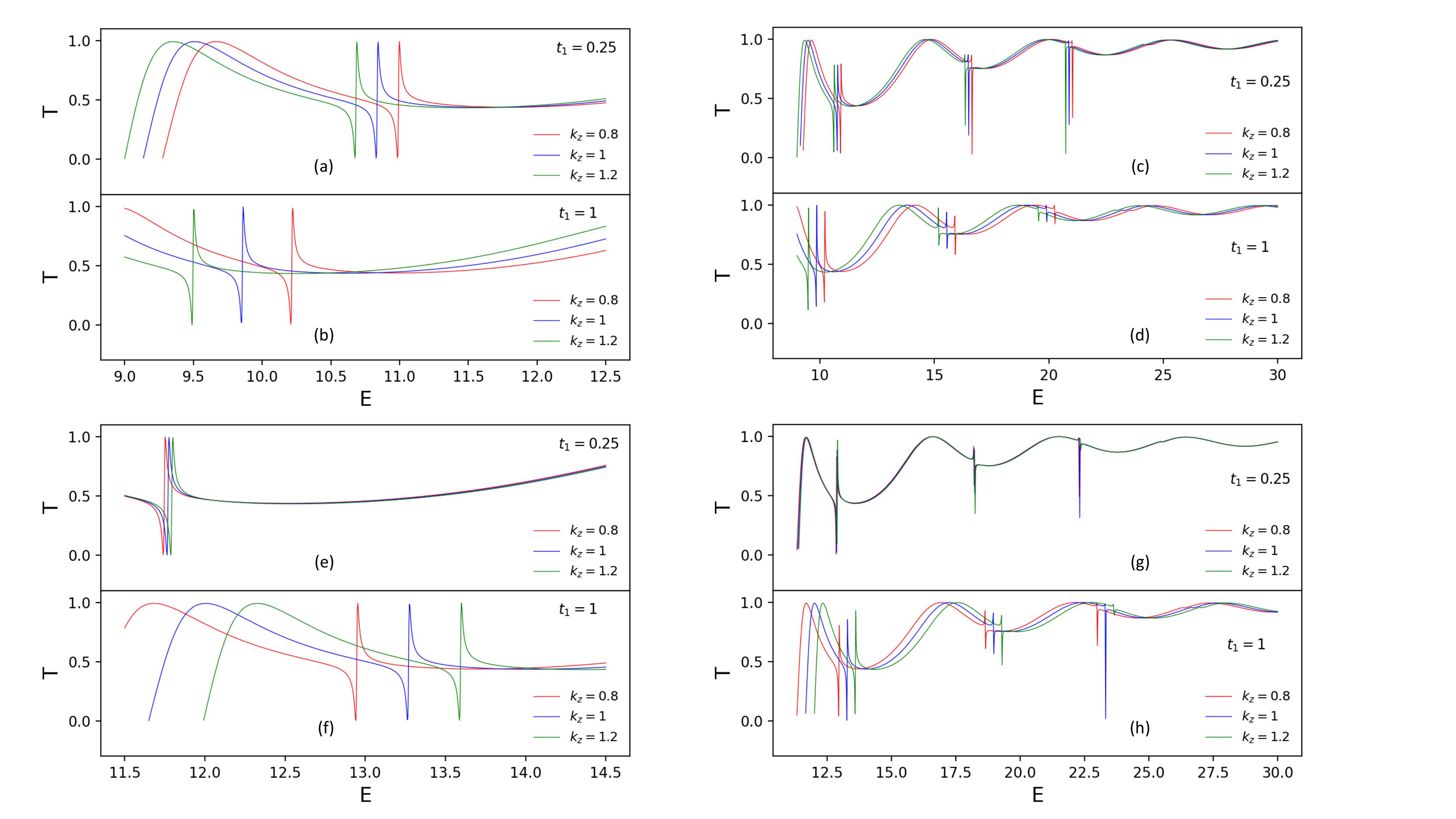}
			
		\caption{Transmission spectra $T$, computed from Eq. (\ref{eq:transmission}), as a function of incident energy $E$ for positive chirality Weyl point with $s=+1$ are shown above keeping $k_y$ fixed while changing value of $k_z$. Here (a,b) display the first Fano resonance and (c,d) show a large energy region of incident energy for $\phi_1=\pi/4$ and $\phi_2=\pi/2$. We repeat (a,b) and (c,d) in (e,f) and (g,h), respectively for $\phi_1=\pi$ and $\phi_2=\pi/2$. Parameters are taken as follows $L=0.6$, $k_y=10$, $V_0=100$, $V_1=1$ and $\omega=15$.}
		\label{fig:3}	
	\end{figure*}

We now proceed with the numerical evaluation of the transmission coefficient ($T$) and reflection coefficient ($R$) from the $S$-matrix. In all cases, we ensure that $R+T=1$. 
We investigate the  Fano resonance profile with the incident energy $E$ by keeping $k_z$ fixed and varying $k_y$ in Figs. \ref{fig:1} and \ref{fig:2} for positive 
and negative chiralities, respectively.  We further explore the Fano resonance characteristics 
by keeping $k_y$ fixed and varying $k_z$ in Figs. \ref{fig:3} and \ref{fig:4} for positive 
and negative chiralities, respectively. These investigations help us understand the tilt-mediated  Fano resonance profile along the direction of tilt momentum and perpendicular to it. 
In all the above scenarios,  we consider two sets of values $(\phi_1,\phi_2) = (\pi/4, \pi/2)$ and $(\phi_1,\phi_2) = (\pi, \pi/2)$ such that effect of the tilt is extensively analyzed.   For our numerical computation, we choose $k_0=\pi/2$, $t=t_z=1$ and $s=\pm 1$, without the loss of generality. For $\phi_1=\pi/4$ and $\phi_2=\pi/2$, we obtain $\eta_{\pm}=|\sqrt{2}t_1 \pm 2\sqrt{2}t_2|$. We find $\eta_{\pm}=|4t_2 \mp 2t_1|$ for $\phi_1=\pi$ and $\phi_2=\pi/2$. We now choose appropriate values of $t_1$ and $t_2$ such that we are able to probe all phases of the WSM model. We reiterate that  $\eta_{\pm}<1$  represent type-I WSMs, and $\eta_{\pm}>1$  represents type-II WSMs. Notice that $t_{1,2}$ are key parameters responsible for tilted spectrum even in the absence of  $\phi_{1,2}$, i.e., $\phi_{1,2}=0$.

\textcolor{black}{The static potential well does not yield any Floquet side bands as there are no additional degrees of freedom that result in energy replication. 
In  the case of the periodically driven potential well, there exists an additional degree of freedom namely, time period or frequency yielding the energy replication in terms of the frequency \cite{F1}. The Floquet theory allows us to probe the time-dependent problem  in a time-independent manner at stroboscopic time intervals after a complete time period. However, the time-independent Floquet Hamiltonian after a full period contains all the Fourier modes in the frequency space which enlarges the dimensionality of the problem by taking into account Floquet side bands.  
One can think of Floquet theory as the temporal analog of the Bloch theory where the spatially periodic array of  potential barriers produces a periodic Bloch wave-function. Note that time period $T$ (spatial periodicity of potential barrier $a$) introduces the repetitive nature of energy (wave-function) with  frequency $\omega$ (wave-vector $k$) in the Floquet (Bloch) theory. The solution of the time-dependent Schr\"odinger equation  takes the form $\psi(t)=e^{i \mu t} \phi(t)$ where quasi-states are periodic $\phi(t)=\phi(t+T)$ and $\mu t$ is defined modulo $2\pi m$ with $m$ being an integer. This further guarantees that quasi-energy $\mu$ is only well-defined up to the driving frequency $\omega$ at stroboscopic time intervals after a complete time period such that $\mu \in [-\omega/2,-\omega/2]$ and $\mu +m\omega$ is an equally valid quasi-energy solution.  The finite values of $m\ne 0$ corresponds to the Floquet side bands.
Connecting with the Bloch theorem, a generic time-independent wave-function can be written as $\psi(r)=e^{i k r} \phi(r)$ with $\phi(r+a)=\phi(r)$ and $kr$ is defined modulo $2\pi m$. This causes the momentum $k$ to be defined up to the reciprocal lattice vector $G=2\pi/a$ such that $k=k+mG$. 
One can hence obtain the extended Brillouin zone of $\mu$ in the frequency space for the Floquet case  which is just equivalent to the extended Brillouin zone of the  wave-functions in the momentum space for Bloch theorem. This results in $\mu_m= \mu_0 + m \omega $ which $m$ represents the multiple Floquet side bands.}

The harmonically driven potential well leads to the formation of multiple copies of quasi-bound states in the potential region which are absent for the static potential well. 
The multiple copies of the quasi-bound state energies that are separated by the photon energy $\hbar \omega$ from each other can be alternatively thought of as quasi-bound Floquet sidebands where  virtual photon transfer processes take place. Once the  incident energy matches with any of these quasi-bound state energies, one encounters a Fano resonance. More precisely, when the energy difference between the incident beam and the quasi-bound state energy is an integer multiple of $\hbar \omega$, an electron can be absorbed  into the quasi-bound state or an electron can be emitted from the quasi-bound state yielding the peak-dip or dip-peak profile of Fano resonance in transmission spectra. Note that the  quasi-bound states are localized inside the potential region while their energies depend on the details of WSM and static potential depth $V_0$.
The concept of Floquet sidebands is also parallelly applicable to the incident and outgoing beams  where their energies are multi-valued and separated by $\hbar \omega$ from each other. The Fano resonance can be equivalently considered as an outcome when the energy of the incident Floquet sideband matches with the quasi-bound states of the well. The relationship between the Fano resonance energy $E_F$ i.e., the value of incident energy where Fano resonance occurs and the quasi-bound state energy $E_b$ is given by $E_F-E_b= n\hbar \omega$   where $n$ can be positive or negative integer. Once we absorb $n\hbar \omega$ inside $E_b$ ($E_F$), we refer to quasi-bound (incident) Floquet sidebands as described above.  In what follows, we are interested in the distribution of the Fano resonances as we vary the different parameters described above. Our aim is to investigate the nature of $E_b$ extensively by tuning the properties of WSM. It can be presumed that $E_b$ incorporates the contribution from the tilt term of WSM as well.

In all cases, we work with the natural units $\hbar = c = e = k_B = 1$. \textcolor{black}{ This helps us to examine the Fano resonance in a simplified manner without altering the qualitative features.} \textcolor{black}{We choose $t_2=0.1$ for all our calculations.} We consider one Floquet sideband around the zero photon-sector in our numerical  analysis, which is sufficient to achieve numerical accuracy with a finite termination of the infinite sum. Note that  $V_1/\omega \ll 1$ in our analysis.  We consider incident energy, potential barrier height, and frequency all in the unit of  $10^{-3}$ eV. The Fermi velocities $t, t_z$ and tilt strength $t_{1,2}$ are measured in the unit of $10^6$ m/s. The momenta $k_{y,z}$ are considered in units of $10^{-9}$ m$^{-1}$. The length of the potential barrier $L$ is in units of  $10^{-9}$ m.
\textcolor{black}{One can increase $L$ and consider it in units of $L^{-6}$ m but the transmission spectra remain qualitatively unaltered \cite{Z17,FW1,Gregefalk23}.} \textcolor{black}{The above choices of the parameters restrict the low-energy model, described in Eq. (\ref{00}), to have energy in the  meV  regime. This is consistent with the long wave length limit where the lattice spacing $a$ is considered in the units of m such that $k\ll 10^0 {\rm m}^{-1}$. On the other hand, the dimension of the system in the tilt direction along $z$-axis $L_z \sim 1/k_z$   is large compared to the barrier width along the transport direction i.e., $x$-direction: $L_z>L$. We do not need to satisfy such condition along $y$-direction as the tilt is only along $z$-direction. However, more importantly, the choice of $L$ has to be comparable to the meV energy scale that we always satisfy. We further note that the Fano resonances are not limited by a particular choice  of the above unitful parameters. Their quantitative features can be altered, however, qualitative behavior remains unchanged. Therefore, the validity of our findings, presented below, is intact as far as the low-energy model of WSM is concerned.}

	
\begin{table}[ht]
	\centering
	\begin{tabular}{|c|c|c|c|c|c|c|}
		\hline
		$(\phi_1,\phi_2)$ &s &$t_1$ &\multicolumn{3}{|c|}{$1^{st}$ Fano Resonance Energy} &$\eta_s$  \\
		\cline{4-6}
		& & &$k_y=8$ &$k_y=9$ &$k_y=10$ & \\
		
		\hline
		\multirow{6}{*}{$(\pi/4,\pi/2)$}
		&\multirow{3}{*}{1}
		&0.25 &10.54 &10.74 &10.92 &0.6364 \\
		& &1 &9.48 &9.68 &9.86 &1.6971 \\
		\cline{2-7}
		&\multirow{3}{*}{-1}
		&0.25 &11.24 &11.44 &11.62 &0.0707 \\
		& &1 &12.31 &12.51 &12.68 &1.1314 \\
		\hline
		\multirow{6}{*}{$(\pi,\pi/2)$}
		&\multirow{3}{*}{1}
		&0.25 &11.39 &11.60 &11.77 &0.1 \\
		& &1 &12.89 &13.09 &13.27 &1.6 \\
		\cline{2-7}
		&\multirow{3}{*}{-1}
		&0.25 &10.39 &10.49 &10.77 &0.9 \\
		& &1 &8.89 &9.09 &9.27 &2.4 \\
		\hline

	\end{tabular}
	\caption{
	Energies associated with Fano resonances for type I ($\eta_s<1$) and type II ($\eta_s>1$) WSMs for both the chiralities $s=\pm1$  (with $k_z=1$ for case 1).}  
	\label{tab:1}
\end{table}


\subsection{Case 1: Fixed $k_z$ and varying $k_y$}

\begin{figure*}[t]
		
			\centering
			\includegraphics[scale = .55]{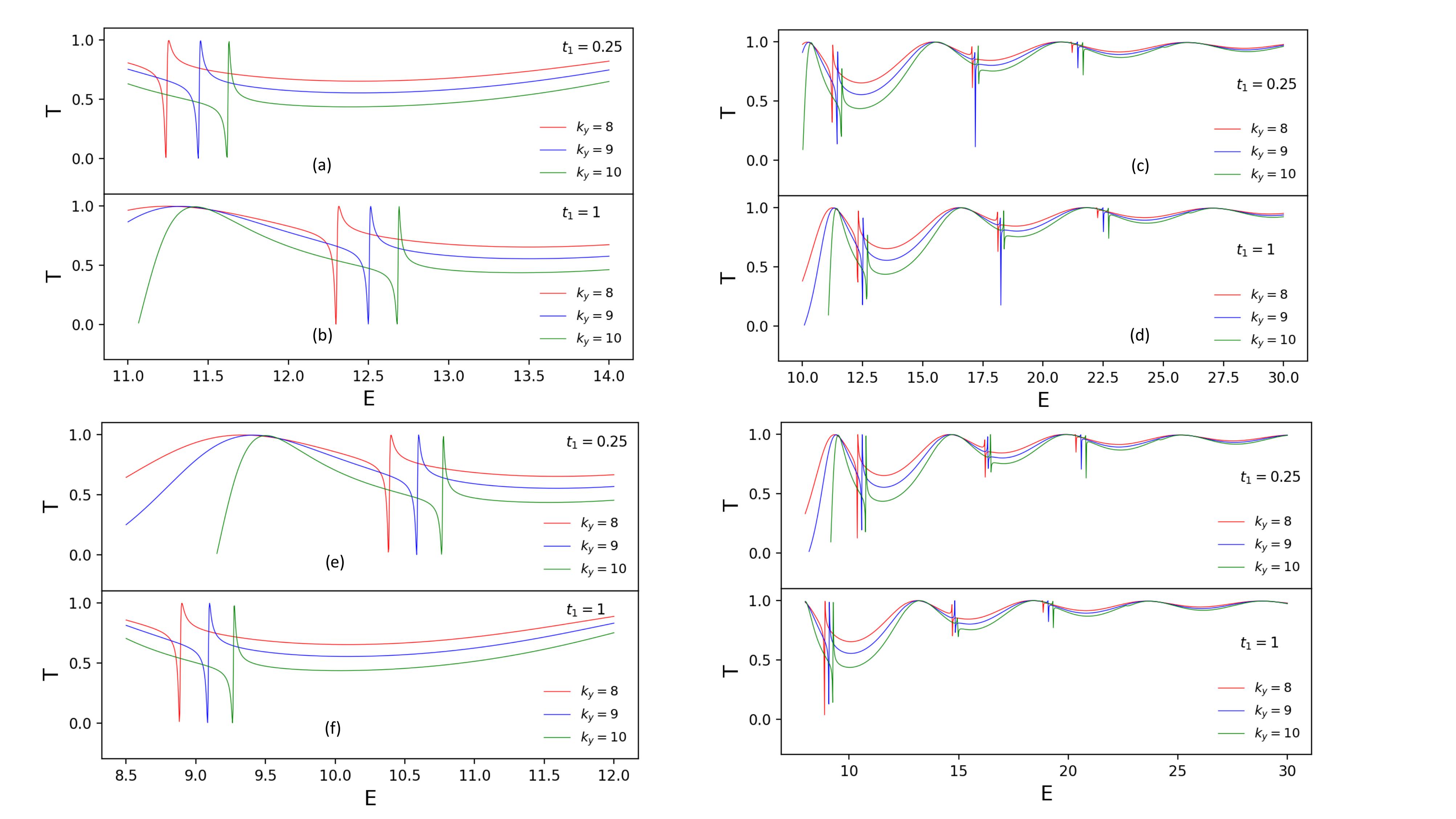}
			
		\caption{ Variation  of $T$ as a function of the  incident energy $E$ for negative  chirality Weyl point with $s=-1$. Fig. \ref{fig:1} is repeated here for  $s = -1$, keeping all other parameters same.}
		\label{fig:2}
	\end{figure*}

We examine the evolution of the transmission spectra in  Figs. \ref{fig:1} and \ref{fig:2} for $s=1$ and $-1$, respectively, and when the tilt factor $t_1$ increases such that we encounter type-I and type-II WSM phases. The details of the  Fano resonance energies are given in table \ref{tab:1}.

The Fano resonances for the larger $k_y$ appear at greater values of incident energy $E$ irrespective of the  choice of the parameters $\phi_{1,2}$,  $t_{1,2}$ and chiralities $s=\pm 1$ as clearly observed by 
scrutinizing the first Fano resonance in  Figs. \ref{fig:1} (a,b,e,f) and \ref{fig:2}  (a,b,e,f). However, we observe significant differences between the cases $s=1$ and $s=-1$ as described below. The Fano resonance energy decreases with $t_1$ for the $s=1$ case and increases with the increase in $t_1$ in $s=-1$ case, for $(\phi_1,\phi_2)=(\pi/4,\pi/2)$. An opposite  trend is noticed for $(\phi_1,\phi_2)=(\pi,\pi/2)$ case. The underlying physics is explained by noting that varying the tilt alters the energies of the  quasi-bound Floquet states, which leads to changes in the energetics
of the Fano resonances - hence they appear at different points as a function of tilt. \textcolor{black}{A complete closed form expression  the  quasi-bound state energy $E_b$ is very useful to understand the emergence of the Fano resonances.} One can naively anticipate the tilt dependence in the above  quantity to be identical to the tilt term in energy of the bare Hamiltonian. This we shall investigate below extensively. 
Passing by, we note that the  quasi-bound state energies for the  untilted and the tilted case differ by the tilt term precisely.

	
Examining the first Fano resonance at energy $E^1_F$ further, we find that 
the Fano resonance energy linearly increases with increasing $k_y$ irrespective of the chirality of the node and tilt strength. This causes the gap between Fano resonance energies, occurring for two different values of $k_y$ to become constant irrespective of the choice of parameters $\phi_{1,2}$,  $t_{1,2}$ and chiralities $s=\pm 1$. This clearly conveys that 
the tilt-mediated part in the quasi-bound state energy is not a function of  $k_y$.  One can decompose the quasi-bound state energy $E_b$ in two parts where the tilt-mediated part $E^t_b$ vanishes if $t_{1,2}=0$ while the untilted normal part  $E^{n}_b$ remains non-zero even if $t_{1,2}=0$; $E_b=E^t_b + E^{n}_b$.
This is qualitatively similar to the static case Eq.~(\ref{eq:energy}) where  $t_{1,2}$ are responsible for the tilt even in the absence of $\phi_{1,2}$.
Importantly, the $k_y$-dependency comes solely from the untilted part of the quasi-bound state energy. This is also similar to the case of the static energy dispersion where the untilted part depends only on $k_y$ while the tilted part is independent of $k_y$.

Interestingly, the  tilt-mediated part $E^t_b$ in the quasi-bound state energy is  a function of $t_{1,2}$, $\phi_{1,2}$ and chirality leading to a change in quasi-bound state energy with all the above parameters while $k_y$ is still kept fixed. This causes  the Fano resonance energies to shift with all the above parameters [see Figs. \ref{fig:1} (a,b,e,f) and \ref{fig:2}  (a,b,e,f)]. One can compute the tilt term as $2 k_z(-s t_1 \cos \phi_1 - 2 t_2 \sin \phi_2)$, provided $k_0=\pi/2$, from the static energy dispersion for all the above parameters to find an quantitative agreement with the shift in the Fano resonance energy. 
For example, the tilt term in the static case becomes more negative [positive]  when $t_1$ increases for $(\phi_1,\phi_2)=(\pi/4,\pi/2)$ [$(\pi,\pi/2)$] and $s=+1$. On the other hand, the untilted part of quasi-bound state energy  remains unaltered as it is independent of 
the tilt parameters $t_{1,2}$, $\phi_{1,2}$. Therefore, the quasi-bound state energy will be shifted by the exact amount as the tilted part of the static energy, demonstrated in Fig. \ref{fig:1} (a,b,e,f). 
This correlation can be further verified for $s=-1$ with $(\phi_1,\phi_2)=(\pi/4,\pi/2)$ [$(\pi,\pi/2)$] where static tilt term increases in the positive [negative] energy side  causing the Fano resonance energy to increase [decrease] with $t_1$ [see Fig. \ref{fig:2} (a,b,e,f)].  One could expect all the above findings to be qualitatively remained if $t_2$ is varied while keeping $t_1$ fixed as the tilt term involves $t_1$ and $t_2$  both. Therefore, the tilt-mediated part of the quasi-bound energy states is exactly given by the static counterpart i.e.,  
$E^t_b=2 k_z(-s t_1 \cos \phi_1 - 2 t_2 \sin \phi_2)$
with $k_0=\pi/2$, while the untilted part $E^n_b$ is \textcolor{black}{anticipated to be a function of the barrier height $V_0$, barrier width $L$, momenta $k_{y,z}$, and Fermi velocities $t,  t_z$. A closed form expression, derived in Appendix. \ref{app3}, suggests the lowering of the barrier height independent part of $E^n_b$ as $L$ increases.   }

We depict the follow-up Fano resonances, namely higher  Fano resonances appearing at higher incident energies $E^n_F$ with $n>1$ after the first Fano resonance, for both the chiralities in 
Figs. \ref{fig:1} (c,d,g,h) and \ref{fig:2}  (c,d,g,h). We notice that the energy gap between two consecutive Fano resonances always decreases for a given value of $k_y$ when the incident energy becomes greater. This is essentially caused by the fact that the quasi-bound state  energies  are not equispaced, rather the energy gap between them decreases. These quasi-bound state energy levels inside the potential well take part in the virtual photon transfer process with the incident beam. The dip-peak structure in the following Fano resonances can be altered as compared to the first Fano resonance. 
We note that since there is no Fermi surface based phenomena in this scattering analysis that even if we increase the tilt strength, our results do not distinguish between the type-I and type-II phases of WSM.  The underlying physics is governed by the tilt altering the energies of the quasi-bound Floquet 
 states, which leads to the changes in the energetics of the Fano resonances.  As a result, they appear at different points as a function of tilt.

	
\subsection{Case 2: Fixed $k_y$ and varying $k_z$}

\begin{figure*}[t]
		
			\centering
			\includegraphics[scale = .55]{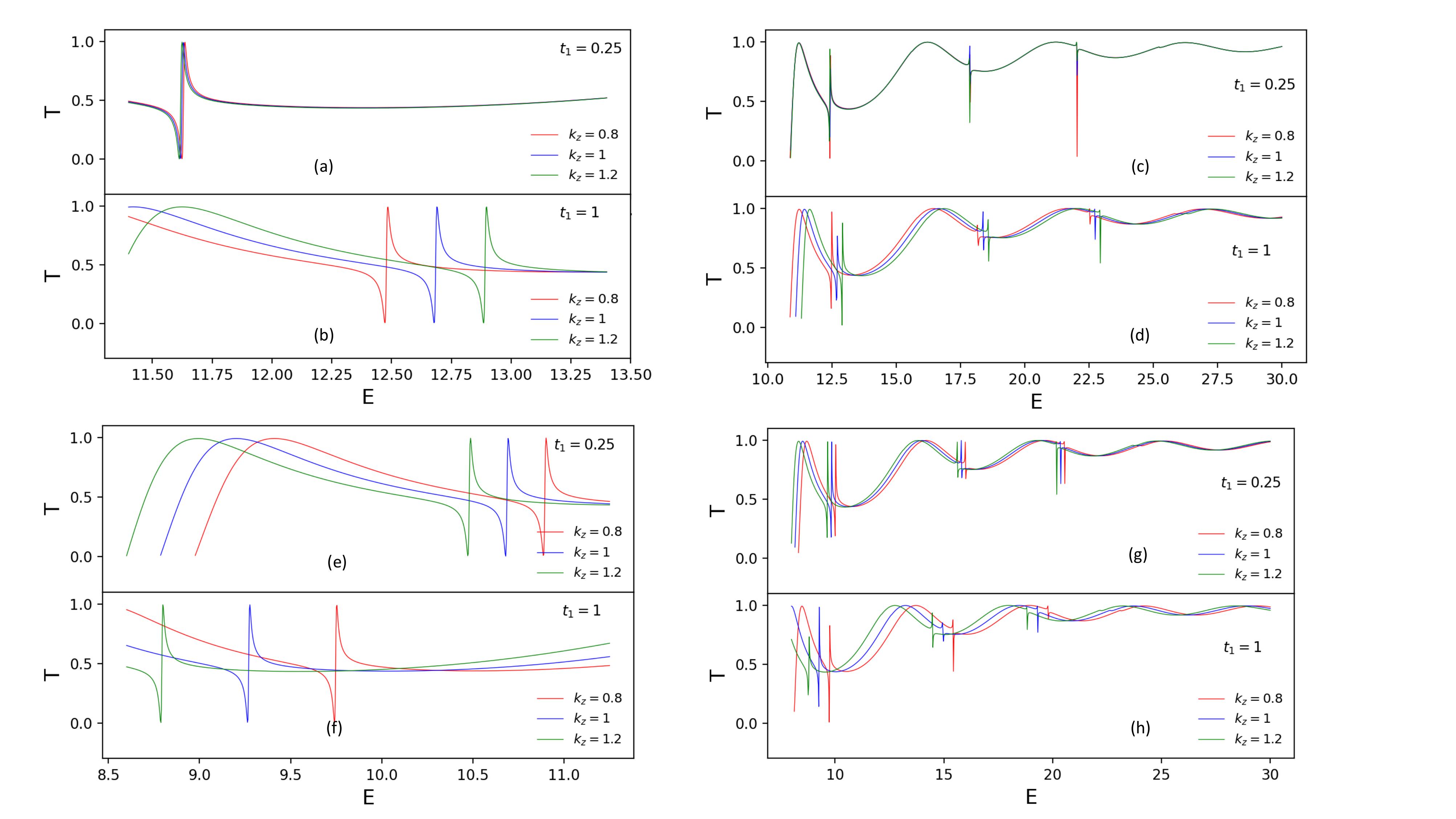}

		\caption{We repeat Fig. \ref{fig:3} for negative chirality Weyl point with $s=-1$.}
		\label{fig:4}
	\end{figure*}

Similar to the earlier subsection, 
we examine the evolution of the transmission spectra in  Figs. \ref{fig:3} and \ref{fig:4} for $s=1$ and $-1$, respectively, when the tilt factor $t_1$ increases. The  Fano resonance energies   are listed in table \ref{tab:2}. Having examined the tilt-mediated effect in $T$ by varying $k_y$, we here further scrutinize the explicit form of $E^t_b$, as already predicted earlier, with respect to  $k_z$.


\begin{table}[H]
	\centering
	\begin{tabular}{|c|c|c|c|c|c|c|}
		\hline
		$(\phi_1,\phi_2)$ &s &$t_1$ &\multicolumn{3}{|c|}{$1^{st}$ Fano Resonance Energy} &$\eta_s$  \\
		\cline{4-6}
		& & &$k_z=0.8$ &$k_z=1$ &$k_z=1.2$ & \\
		
		\hline
		\multirow{6}{*}{$(\pi/4,\pi/2)$}
		&\multirow{3}{*}{1}
		&0.25 &11.06 &10.92 &10.77 &0.6364 \\
		& &1 &10.21 &9.85 &9.49 &1.6971 \\
		\cline{2-7}
		&\multirow{3}{*}{-1}
		&0.25 &11.63 &11.62 &11.61 &0.0707 \\
		& &1 &12.48 &12.68 &12.89 &1.1314 \\
		\hline
		\multirow{6}{*}{$(\pi,\pi/2)$}
		&\multirow{3}{*}{1}
		&0.25 &11.75 &11.77 &11.79 &0.1 \\
		& &1 &12.94 &13.26 &13.59 &1.6 \\
		\cline{2-7}
		&\multirow{3}{*}{-1}
		&0.25 &10.93 &10.76 &10.56 &0.9 \\
		& &1 &9.74 &9.26 &8.79 &2.4 \\
		\hline

	\end{tabular}
	\caption{
	Energies associated with Fano resonances for type I ($\eta_s<1$) and type II ($\eta_s>1$)  WSMs for both the chiralities $s=\pm1$  (with $k_y=10$ for case 2).} 
	\label{tab:2}
\end{table}

The first Fano resonance energy $E^1_F$, corresponding to positive chirality, decreases [increases] with increasing $t_1$ with a given value of $k_z$ for $(\phi_1,\phi_2)=(\pi/4,\pi/2)$ [$(\pi,\pi/2)$] as depicted in Figs.~\ref{fig:3} (a,b,e,f).  The situation is reversed for negative chirality as shown in 
Figs.~\ref{fig:4} (a,b,e,f). This feature is also observed in the previous case. On the other hand, the marked difference for the present case is the following: 
the energy gap between two successive Fano resonances, associated with two different values of the tilt momentum $k_z$, varies with increasing $t_1$. This energy gap also changes if one changes the chirality as well as $\phi_{1,2}$.  This suggests that the tilt-mediated part $E^t_b$ in the quasi-bound state energy depends on
$k_z$, $t_{1,2}$, $\phi_{1,2}$ and $s$. This can be quantitatively understood from the tilted part in the static energy dispersion Eq.~(\ref{eq:energy}).
As discussed earlier, 
the untilted part $E^n_b$ of the quasi-bound state energy is independent of $t_{1,2}$, $\phi_{1,2}$ and $s$. The tilt momentum $k_z$  appears both in the tilted as well as untilted part of the above energy.  
Therefore, once we vary all the parameters $t_{1,2}$, and $\phi_{1,2}$ including $k_z$ for a given chirality, both the untilted and tilted-mediated parts in the quasi-bound state energy change in a non-trivial manner.


	
The above phenomena lead to a rich Fano resonance profile that is an admixture of effects coming from tilt factors $t_{1,2}$ and tilt momentum $k_z$ simultaneously. This can be further understood from the fact that beyond a certain value of $t_1$ the order of the Fano resonance with $k_z$ changes i.e., quasi-bound state energies increase 
(decrease) with increasing $k_z$ instead of   decrease (increase). This is noticed in Fig.~\ref{fig:4} (a,b).  
Therefore, the  appearance of Fano resonances with $k_z$ is relative to the choice of the parameters such as $t_{1,2}$, and $\phi_{1,2}$. 
One could expect  similar findings  if $t_2$ is varied while keeping $t_1$ fixed as the tilt term involves $t_1$ and $t_2$  both.         
A careful investigation suggests that the Fano resonance energy changes linearly with increasing $k_z (t_1)$ for a given value of $t_1 (k_z)$. This causes the energy separation between two subsequent Fano resonances to vary linearly with increasing the product of $t_1$ and  $k_z$. This collective behavior indicates that tilt-mediated part of the quasi-bound state energies 
behave in a qualitatively different manner as compared to the untilted counterpart. 

We now turn our attention to the follow-up Fano resonance at  energies $E^n_F$ with $n>1$ profiles when the incident energy increases as shown in Figs.~\ref{fig:3} (c,d,g,h) and \ref{fig:4} (c,d,g,h) for positive and negative chiralities, respectively. Similar to the earlier section, we find that the energy gap between two successive higher Fano resonances   decreases with increasing the incident energy for a given value of  $k_z$. Such repeated occurrence of Fano resonances with a variation in the dip-peak structures is commonly visible when $t_1$ increases as expected. The existence of multiple quasi-bound states with unequal energy spacing can be inferred as well from the above findings. 
		
\section{Discussion}
\label{sec5}

Previously we discussed the explicit form of $E^t_b$ for a given value of $t_1$ while varying either collinear momentum $k_z$ or transverse momentum $k_y$. In this section, we show that the specific form of $E^t_b$ predicted from the earlier discussion is not special with respect to the values of $t_1$ chosen earlier. The expression of $E^t_b=2 k_z(-s t_1 \cos \phi_1 - 2 t_2 \sin \phi_2)$
with $k_0=\pi/2$ is generic and not limited to a special range of $t_{1,2}$ and $\phi_{1,2}$. In order to strengthen our findings on the tilt-mediated part of quasi-bound state energy, we examine the evolution of Fano resonance energy $E_F$ with $t_1$ as described below.

\begin{figure*} [t]
					
						\centering
						\includegraphics[scale = .5]{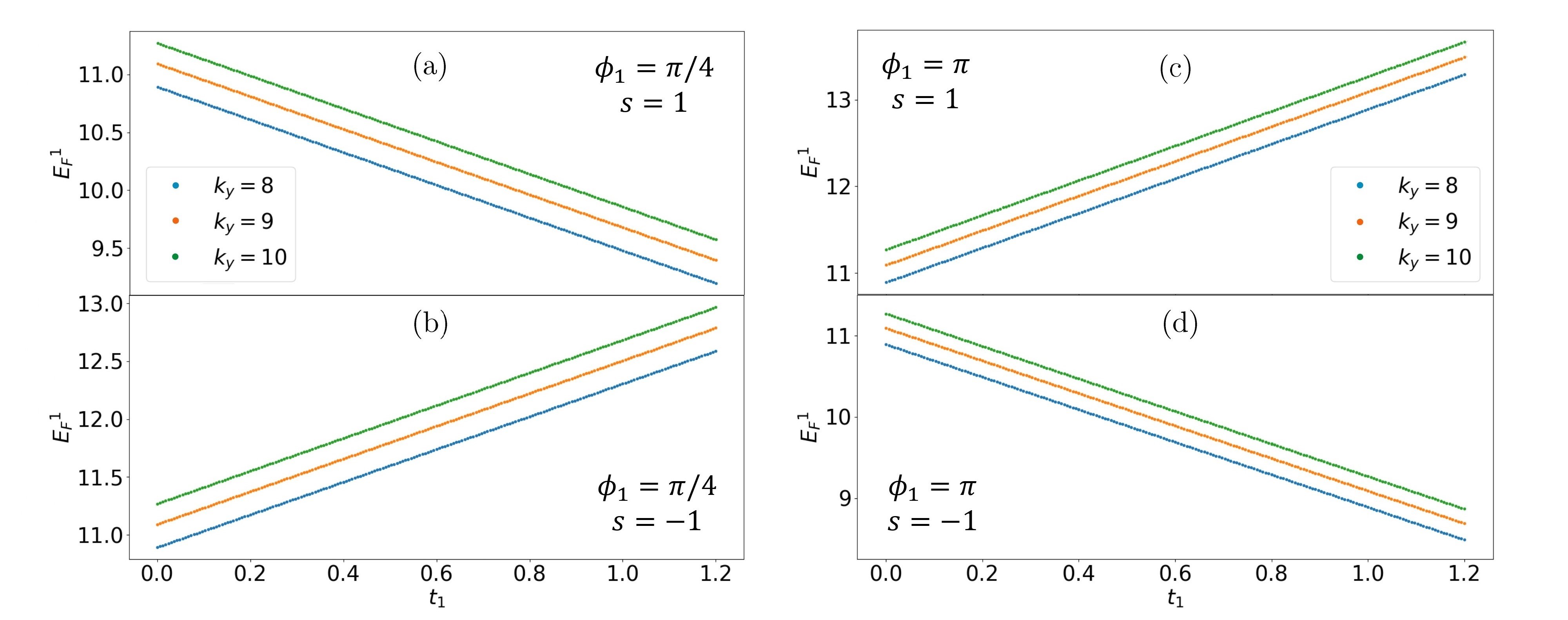}
						
				\caption{The energy $E^1_F$ corresponding to first Fano resonance
				as found in Figs. \ref{fig:1} and \ref{fig:2},
				are shown as a function of $t_1$. We consider $(\phi_1, \phi_2) =(\pi/4, \pi/2)$ and   $(\pi, \pi/2)$ for (a,b) and (c,d), respectively. We choose $s=+1$ and $-1$ for (a,c) and (b,d), respectively. Parameters are taken as follows $L=0.6$, $k_z=1$, $V_0=100$, $V_1=1$ and $\omega=15$. }
				\label{fig:5}
			\end{figure*}

\begin{figure*} [t]
					
						\centering
						\includegraphics[scale = .5]{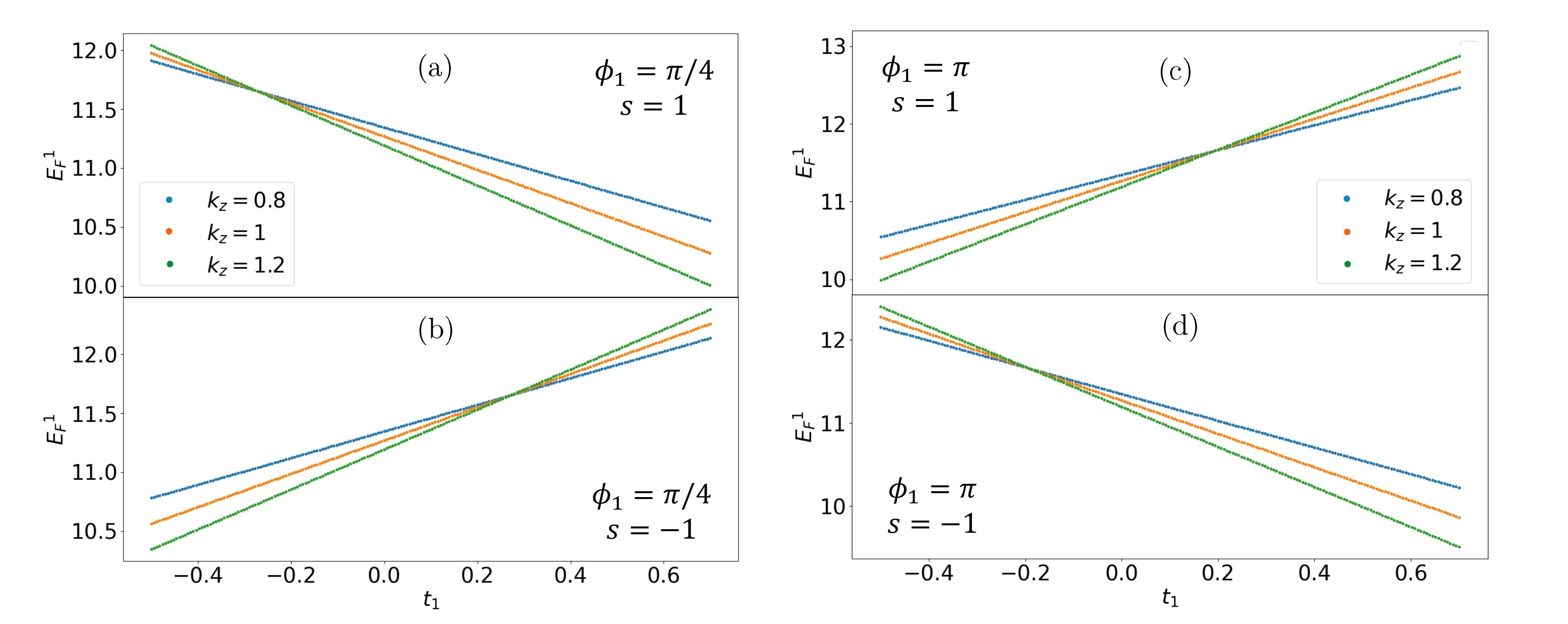}
						
					\caption{The energy $E^1_F$ corresponding to first Fano resonance as found in Figs. \ref{fig:3} and \ref{fig:4},
					are shown as a function of $t_1$. We consider $(\phi_1, \phi_2) =(\pi, \pi/2)$ and   $(\pi/4, \pi/2)$ for (a,b) and (c,d), respectively. We choose $s=+1$ and $-1$ for (a,c) and (b,d), respectively. Parameters are taken as follows $L=0.6$, $k_y=10$, $V_0=100$, $V_1=1$ and $\omega=15$.}
				\label{fig:6}
			
				\end{figure*}

In particular, we plot the incident energies $E^1_F$, corresponding to  the first Fano resonance, as a function of $t_1$
keeping $k_y$ as a parameter (see Fig.~\ref{fig:5}).    We investigate the case of $(\phi_1,\phi_2)=(\pi/4,\pi/2)$ [$(\pi,\pi/2)$] in Fig.~\ref{fig:5} (a,b) [(c,d)]. We repeat Fig.~\ref{fig:5} with $k_z$ as a parameter in Fig.~\ref{fig:6}. The linear behavior of Fano resonance energy with $t_1$ infers that the  quasi-bound state energy varies linearly with $t_1$.


				
This further boils down to the fact that the tilt-mediated  part  shows linear dependence with $t_1$.   The parallel straight lines, on the other hand,  for different values of $k_y$ suggests that the quasi-bound state energies change in an identical manner with $k_y$ irrespective of the values of $t_1$. This is in accordance with the fact that tilt-mediated part $E^t_b$ of quasi-bound state energy does not depend on $k_y$. Interestingly, the untilted part of the above energy is expected to vary linearly with $k_y$ as evident from the equidistant parallel lines. The parallel nature of the straight lines confirms that the energies of the quasi-bound state do not involve any product term $t_1 k_y$, rather it  depends on $k_y$ and $t_1$ separately.

The positive and negative slopes of these parallel lines denote the coefficient in front of these energies that depend on chirality as well as the choice of $\phi_{1,2}$.  We find the sign of the coefficient reverses as chirality switches its sign while $\phi_{1,2}$ are kept fixed.  The change in the sign of chirality  $s =1$ to $s=-1$ can alter  the sign of the slopes of these straight lines without affecting their parallel nature.  This further suggests that energies depend on the product of $s t_1$ while $k_y$  does not couple with $s$. One can understand all of the above findings by analyzing the  tilt term $2 k_z(-s t_1 \cos \phi_1 - 2 t_2 \sin \phi_2)$ in the static energy dispersion   with a given set of parameters.  

Examining Figs. \ref{fig:6}(a,b) and (c,d) for $(\phi_1, \phi_2) =(\pi, \pi/2)$ and   $(\pi/4, \pi/2)$ respectively,
we find that the first Fano resonance energy $E^1_F$ changes linearly with tilt parameter $t_1$ while the slope depends on $k_z$. This results in intersecting straight lines for different values of $k_z$  unlike the previous case of parallel lines in Fig. \ref{fig:5}. On the other hand, these energies also depend linearly on $k_z$  given a fixed value of $t_1$. Interestingly, for certain values of $t_1$,  the Fano resonance energies for different values of $k_z$ become identical. This  equi-energy point changes to a different value of $t_1$ once $t_2$ and/or the chirality $s$ changes for a fixed parameter set of $\phi_{1,2}$. 
This refers to the fact that there exists an additional $t_2 k_z$ tilt term in the energy, and can be clearly understood from the tilt term $2 k_z(-s t_1 \cos \phi_1 - 2 t_2 \sin \phi_2)$ of the static  model. The intersection point is observed when the tilt-mediated part of the quasi-bound state energy 
vanishes i.e., $E^t_b=0$  for a certain set of $t_1^*$ and $t_2^*$ values given $\phi_{1,2}$ fixed. The values of $t_{1,2}^*$, such that $-s t^*_1 \cos \phi_1 = 2 t^*_2 \sin \phi_2$, can be exactly computed from the tilt term in the static model. 

				
The non-parallel intersecting nature of the straight lines confirms that energies of the quasi-bound state  involve the product term $s t_1 k_z$. This collective dependence rules out the individual dependence on $k_z$ and $t_1$. The change in the sign of chirality  $s =1$ to $s=-1$ alters the sign of the slopes of these straight lines without affecting the intersecting  nature.
The energies change linearly with $k_z$ while the proportionality factor depends on the product of  $ s t_1$ and $ t_2$ separately as evident from the tilt term in the low-energy Hamiltonian Eq. (\ref{00}).  This describes the  intersecting nature of the plots  and alteration in order of the Fano resonance peaks with $k_z$ upon changing $t_1$ across $t_1^*$ keeping all the other parameter fixed, as noticed in Fig.~\ref{fig:4} (a,b).

\textcolor{black}{Having discussed the parameter dependent response in Figs. \ref{fig:5} and \ref{fig:6}, we here analyze the intersecting nature of $E^1_F$ profiles with respect to the tilt orientation of Weyl points. Once the tilt term changes from positive to negative values the  intersecting point in  $t_1$ can be observed when $E^1_F$ curves for different values of $k_z$ meet. The order of $E^1_F$ curves for different $k_z$  gets reversed once the intersecting point is met at   
$-s t^*_1 \cos \phi_1 = 2 t^*_2 \sin \phi_2$ irrespective of the chirality. One interesting observation is that for $|t_1|<  |2 t_2 \sin \phi_2/\cos \phi_1|$,  both chirality Weyl points share identical tilt orientaiton profiles. This is evident from the identical nature of the order of $E^1_F$ curves associated with $k_z$ for both $s=\pm 1$ in \ref{fig:6} (a,b) and (c,d) with $(\phi_1,\phi_2)=(\pi/4,\pi/2)$ and $(\pi,\pi/2)$, respectively. On the other hand, for $|t_1|>  |2 t_2 \sin \phi_2/\cos \phi_1|$, opposite chirality Weyl points show opposite tilt orientation profiles as the orders of $E^1_F$ curves get reversed with $k_z$ outside the window of $-2 t_2 \sin \phi_2<t_1 \cos \phi_1<  2 t_2 \sin \phi_2$ for $s=\pm 1$. Therefore,   
the choice of the tilt term, considered here, allows us to have two opposite tilted Weyl points in a certain parameter regime.}

\textcolor{black}{This results in the fact that  without altering the tilt term manually, we can 
probe the Fano resonance energy, associated with opposite tilted Weyl points, with tilt strength. 
This is markedly different from a simple tilt term, governed by a single set of parameter i.e., either by $(t_1,\phi_1)$ or $(t_2,\phi_2)$, that does not necessarily yield two different tilt orientaitons among two opposite chirality Weyl points. A simplified tilt term with tilt momentum $k_z$ and one set of parameters ($t_1,\phi_1$) can also partially mimic the same results presented in Figs. \ref{fig:5} and \ref{fig:6} when the tilt orientations of  Weyl points are manually made opposite to each other. 
Our choice of the tilt term is solely responsible for the interesting nature of $E^1_F$ lines for different values of the tilt momentum $k_z$. The tilt orientations for both the opposite chirality Weyl points can be simultaneously controlled by the two sets of parameters  $(t_1,\phi_1)$ and $(t_2,\phi_2)$ and hence it does not require any manual tuning. This leaves recognizable change in the Fano resonance profile with tilt strengths while adopting a generalized form of the tilt term. In other words, the orientation of tilt-driven intersecting points and their distinct locations in terms of tilt strength for different chirality are the key signatures of the particular tilt term considered here. 
Importantly,  by varying $t_1$ it is possible to obtain the type-I,  type-II and hybrid phase where one Weyl point is type-I and its chiral counterpart is type-II \cite{Nag22}. However, distinct signature of the above phases may not be obvious from $E^1_F$ lines. Nevertheless, the tilt term, considered here, has all these ingredients leading to  tilt orientation mediated distinct transport response for opposite chirality Weyl points. Our findings are thus more general compared to that of a conventional tilt term.  }

\textcolor{black}{The interference between the open (continuum states) and closed (Floquet sidebands) channels leads to Fano resonance 
where bound state energy plays a crucial role. The bound state energy is not simply given by the energy dispersion of the Weyl semimetal rather it is given by 
\begin{equation}
 E_b = \left[V_{0} \pm  \sqrt{(2 l+1)^{2} \frac{\pi^{2} t^{2}}{4 L^{2}}+t^{2} k_{y}^{2}+ t_{z}^{2}k_{z}^{2} \sin ^{2} k_{0} }\right]+ k_{z} \tilde{t}
 \label{eq_Eb}
\end{equation}
with $\tilde{t}=2t_1\sin(\phi_1-sk_0 )+4t_2 \sin(\phi_2 -2sk_0)$ and $l=0,\pm 1, \pm 2, \cdots$.
The barrier width and height are also important to determine the bound state energy.
The tilt-independent part is thus given by $E^n_b=V_{0} \pm  \sqrt{(2 l+1)^{2} \frac{\pi^{2} t^{2}}{4 L^{2}}+t^{2} k_{y}^{2}+ t_{z}^{2}k_{z}^{2} \sin ^{2} k_{0} }$ while the tilt-mediated part is $E^t_b=k_{z} \tilde{t}$ as predicted by our prior analysis with $k_0=\pi/2$.
The detailed derivation on S-matrix formulation is given in Appendix \ref{app3}.
When the incident energy $E$ matches with the bound state of the Floquet sidebands $E_b + n \omega$ (with $\hbar=1$, and $n=0,\pm 1, \pm 2, \cdots$), we find Fano resonance at $E=E_F=E_b + n \omega$. Note that barrier height $V_0$ is absorbed within $E_b$. This is the reason that a simple kinematic relation between $E_F$ and the energy dispersion, given by $E_F+n\omega=\tilde{E}(q_x,k_y,k_z,V_0)$ can describe the emergence of Fano resonance \cite{Fano1}.  Note that $\tilde{E}(q_x,k_y,k_z,V_0)$ is not exactly given by the bulk energy dispersion in the presence of the static potential well which is found to be $\tilde{E}_{s,u}=E_{s,u}+V_0$ with $E_{s,u} = \tilde {t} k_z + u\sqrt{t^2(k_x^2 + k_y^2) + t_z^2k_z^2 \sin^2 k_0}$ and $u=\pm$. The potential effectively modifies $k_x$ to $q_x$ as $q_x=2\pi m/L$ in $E_b$ while the other terms inside the square root are proportional to $k^2_{y,z}$ and are exactly the same as that of the bulk energy dispersion. The tilt term appears in an identical way for $E_b$ and  $\tilde{E}_{s,u}$. Therefore, $\tilde{E}(q_x,k_y,k_z,V_0)$ can only be regarded as $\tilde{E}_{s,u}$ if $k_x\equiv q_x=2\pi m/L$. The kinematic relation essentially dictates the 
shapes of the transmission of spectra at Fano resonances that are universal either exhibiting a peak followed by a dip or vice versa irrespective of the details of the underlying model. The increase (decrease) in the barrier length $L$ reduces (enhances) the energy of the quasi-bound state resulting in  the shift of Fano resonances as incident energy $E$ varies.}

\textcolor{black}{We now discuss the impact of the 
underlying three-dimensional dispersion of WSM  on the transmission spectra. In addition to the one(two)-dimensional problem, here we have  momenta $k_{y,z}$ transverse to the propagation direction along $x$-axis. As a result, we have better tunability on the location of Fano resonance while their profiles are studied in Figs. \ref{fig:1}, \ref{fig:2}, \ref{fig:3}, and \ref{fig:4} by varying $k_y$ and $k_z$ in a respective manner. To be precise, the  parallel (intersecting) nature of Fano resonance energy with tilt strength $t_1$  in Fig. \ref{fig:5} (Fig. \ref{fig:6}) is due to the fact that tilt momentum is unaltered  (varied). These features are not very obvious from the earlier one- and/or two-dimensional studies \cite{Fano1,Z16} revealing an intriguing manifestation of the tilt in the energy dispersion. }

\textcolor{black}{It is difficult to differentiate between the  type-I/under-tilted and type-II/over-tilted WSM as Fermi surface physics is not captured from our analysis. In order to apprehend the change in  the shape of Fermi pocket into the transmission spectra, one has to invoke the concept of chemical potential. In addition to the incident energy $E$,  the bias voltage between the left and right sides of the potential well  $\mu_L$ and $\mu_R$ can be considered to investigate the distinct transmission profiles caused by point-like and pocket-like Fermi surface. The parallel (intersecting) nature of the Fano resonance energy with tilt strength in Fig. \ref{fig:5} (Fig. \ref{fig:6}) can be altered for different choices of gate voltage in the presence of leads.  The variation with chemical potential i.e., gate voltage can  serve as another possible route to differentiate between type-I and type-II WSM. This is a limitation of our present work that  we leave open for future exploration by considering an appropriate Landauer-B\"uttiker version of the three-dimensional lattice Hamiltonian of WSM. }

\textcolor{black}{The Fano resonance spectral profile is characterized by energy shift $F$, spectral width $\Gamma$  and an asymmetry parameter $q$ \cite{Fano61,Shore67,Smith60}. The first two terms of $F$ and $\Gamma$ are real and imaginary parts, respectively, of a self-energy of the scattering system in question. 
The coupling between the open and closed channels, lying outside and inside the potential regions respectively, can be regarded as  $V_E$. To be precise, $V_E$ is represented by $V_1$ multiplied by an overlap between the associated discrete and continuum wavefunctions in the $x$-direction. importantly,  $q$ depend on $|V_E|^2$ and a phase factor of $V_E$. The Fano peak position is associated with the extremum of the first derivative of the scattering phase shift with respect to $E$. This is related to the time-delay matrix as obtained from scattering theory.  One can 
anticipate the spectral width $\Gamma$ of the Fano peak from the above the derivative rule. One can also obtain $q$ qualitatively by fitting the appropriate formula with regard to the relevant channels \cite{Fano61}. It would be interesting to study the energy shift $F$ and asymmetry parameter $q$ from the scattering matrix formalism in future. However, we can comment that the sharpness of the resonance structure is affected by the tilt term as noticed in Figs. \ref{fig:1}, \ref{fig:2}, \ref{fig:3}, and \ref{fig:4}. }

Controlling material parameters in general is considered tricky in the experimental community as the tilt can easily be intrinsically present in Dirac semimetals and WSMs  with linear as well non-linear band dispersion \cite{Nag22}. We, however, point out a few ways in which our results may be tested in a laboratory. Given the fact that  uniaxial strain is found to be useful to introduce tilt in the energy dispersion of graphene \cite{Milicevic19},  applying strain can also cause a variation in the tilt of the WSMs. This method relies on altering the spatial and temporal components of the gauge field \cite{40,55,56} and maybe used to find Fano resonances when driven by a laser field. Another way in which tilt variations appear is through periodic driving, which has been reported to produce type-I and
type-II WSMs \cite{57}. Notice that the tilted Dirac dispersion has been predicted to appear in two-dimension for quinoid-type \cite{Goerbig08}  and hydrogenated graphene \cite{Lu16},  8-Pmmn Borophene \cite{Zabolotskiy16}, planar arrays of carbon nanotubes \cite{Polozkov19}. The tilted Dirac cones  have been experimentally observed in photonic Lieb-kagome lattices \cite{Lang23}.  We believe that  
the above theoretical predictions and experimental techniques will be instrumental in future to engineer the tilted dispersion of WSMs in three-dimension.


\section{Conclusion}
\label{sec6}

In this work, we consider harmonically driven quantum well to study the Fano resonances in tilted WSMs by investigating the transmission spectra using the Floquet scattering theory technique. The potential is oriented in the transverse direction with respect to the tilt allowing us to study the transmission by varying momentum along and perpendicular to the tilt direction. Our study sheds light on the effect of tilt on the Fano resonance where the Fano resonance energy changes linearly with tilt strength. This is caused by the tilt-mediated part of the energy associated with the quasi-bound state residing inside the potential well. For fixed value of the collinear  (transverse) momentum with respect to the tilt direction, the energy difference between two Fano resonances, appearing for two adjacent values of transverse (collinear) momentum, does not (does) change with tilt strength. This finding clearly indicates that 
collinear (transverse) momentum is coupled (decoupled) with tilt strength in the tilt-mediated part of the quasi-bound state energy. The variation of the first Fano resonance energy with tilt strength further confirms that the tilt-mediated part of the above energy is the same as the tilt term present in the static energy dispersion. We investigate the evolution of transmission profile with respect to the other tilt parameters and chirality of the Weyl points extensively, from which we exact form of the tilt-mediated part is further verified.  The normal part of the quasi-energy spectrum is dependent on the collinear as well as the transverse momentum resulting in the rich profile of Fano resonance in the parameter space.

Our work thus uncovers the complex interplay between the tilt and Fano resonance by probing the tilt-mediated part of quasi-bound state energy. \textcolor{black}{while the closed form of the later yields significant insight to understand the numerical results. } \textcolor{black}{The generic form the tilt term allows us to explore the tilt orientation mediated response of the Weyl points that is not usually captured by conventional tilt terms.}
We believe that our study might play an important role in understanding the experimental findings such as currents for periodically driven systems  as far as the microscopic mechanisms are concerned. In real mesoscopic systems, the potential region is confined in a finite region of space and hence edge effects are important. Therefore, it would be interesting to study the edge effects in these systems. At the same time, the dynamics of Fano resonance under magnetic field,  disorder, interaction and other dephasing mechanism might be important for future studies.

\section{Acknowledgement}
We acknowledge the technical help from  Rui Zhu.  
TN would like to thank  Ipsita Mondal for the
initial discussions on the Floquet scattering techniques. TN  also thanks Anton Gregefalk for fruitful discussions.  TN acknowledges NFSG ``NFSG/HYD/2023/H0911" from BITS Pilani.

\appendix

\onecolumngrid

\section{Boundary matching}
\label{app1}

In this appendix, we discuss the boundary conditions in details for the wave-functions to be continuous across the boundary along with their derivations. 
This  allows us to relate the  coefficients $A_n$, $B_n$, $a_m$ and $b_m$ as discussed in Sec.~\ref{sec3}. 

Applying the boundary condition$x=-L/2$ we obtain,
\begin{equation}
    \begin{aligned}
        &A_n^iN_{n_+}e^{-ik_{xn}L/2}+ A_n^0N_{n_-}e^{ik_{xn}L/2}=\sum_{m=-\infty}^{\infty}[a_mN'_{m_+}e^{-iq_mL/2}+b_mN'_{m_-}e^{iq_mL/2}]J_{n-m}\left(\frac{V_1}{\hbar\omega}\right)
    \end{aligned}
\end{equation}

\begin{equation}
\begin{aligned}
     &A_n^iN_{n_+}S_+(E_n)e^{-ik_{xn}L/2}+ A_n^0N_{n_-}S_-(E_n)e^{ik_{xn}L/2}\\
     & = \sum_{m=-\infty}^{\infty}[a_mN'_{m_+}S'_+(E_m)e^{-iq_mL/2}+b_mN'_{m_-}S'_-(E_m)e^{iq_mL/2}]J_{n-m}\left(\frac{V_1}{\hbar\omega}\right)
\end{aligned}
\end{equation}\\

Similarly, boundary conditions at $x=L/2$ yields,
\begin{equation}
    \begin{aligned}
        &B_n^iN_{n_-}e^{-ik_{xn}L/2}+ B_n^0N_{n_+}e^{ik_{xn}L/2}\\
        &=\sum_{m=-\infty}^{\infty}[a_mN'_{m_+}e^{iq_mL/2}+b_mN'_{m_-}e^{-iq_mL/2}]J_{n-m}\left(\frac{V_1}{\hbar\omega}\right)
    \end{aligned}
\end{equation}
\begin{equation}
    \begin{aligned}
    & B_n^iN_{n_-}S_-(E_n)e^{-ik_{xn}L/2}+B_n^0N_{n_+}S'_+(E_n) e^{ik_{xn}L/2}\\
    & = \sum_{m=-\infty}^{\infty}[a_mN'_{m_+}S'_+(E_m)e^{iq_mL/2}+b_mN'_{m_-}S'_-(E_m)e^{-iq_mL/2}]J_{n-m}\left(\frac{V_1}{\hbar\omega}\right)
    \end{aligned}
\end{equation}

Now, from above equations,
\begin{equation}
    \begin{aligned}
        &A_n^i[S_-(E_n)-S_+(E_n)]e^{-ik_{xn}L/2}\\
        &=\frac{1}{N_{n_+}}\sum_{m=-\infty}^{\infty}[a_mN_{m_+}(S_-(E_n)-S'_+(E_m))e^{-iq_mL/2}+b_mN_{m_-}(S_-(E_n)-S'_-(E_m))e^{iq_mL/2}]J_{n-m}\left(\frac{V_1}{\hbar\omega}\right)
    \end{aligned}
\end{equation}
\begin{equation}
    \begin{aligned}
        &B_n^i[S_-(E_n)-S_+(E_n)]e^{-ik_{xn}L/2}\\
        &=\frac{1}{N_{n_-}}\sum_{m=-\infty}^{\infty}[a_mN_{m_+}(S'_+(E_m)-S_+(E_n))e^{iq_mL/2}+b_mN_{m_-}(S'_-(E_m)-S_+(E_n))e^{-iq_mL/2}]J_{n-m}\left(\frac{V_1}{\hbar\omega}\right)
    \end{aligned}
\end{equation}

Now we define the matrices as - 
\begin{equation}
    (M_{sa}^{\pm})_{nm}=N_{m+}^{\prime}[(\frac{1}{N_{n+}}) (S_{+}^{\prime}(E_m)-S_{-}{E_n}) e^{i(k_{xn}-q_m)L/2}\pm(\frac{1}{N_{n-}})(S_{+}{E_n}-S_{+}^{\prime}(E_m))e^{i(k_{xn}+q_m)L/2}]J_{n-m}\left(\frac{V_1}{\hbar\omega}\right)
\end{equation}
\begin{equation}
    (M_{sb}^{\pm})_{nm}=N_{m-}^{\prime}[(\frac{1}{N_{n+}}) (S_{-}^{\prime}(E_m)-S_{-}{E_n}) e^{i(k_{xn}+q_m)L/2}\pm(\frac{1}{N_{n-}})(S_{+}{E_n}-S_{-}^{\prime}(E_m))e^{i(k_{xn}-q_m)L/2}]J_{n-m}\left(\frac{V_1}{\hbar\omega}\right)
\end{equation}
\begin{align}
    (M_{cA}^{1})_{nm}=\frac{N'_{m+}}{N_{n-}}e^{-i(q_{m}+k_{xn})L/2} J_{n-m}\left(\frac{V_1}{\hbar\omega}\right)\\
    (M_{cA}^{2})_{nm}=\frac{N'_{m-}}{N_{n-}}e^{i(q_{m}-k_{xn})L/2} J_{n-m}\left(\frac{V_1}{\hbar\omega}\right)\\
    (M_{cB}^{1})_{nm}=\frac{N'_{m+}}{N_{n+}}e^{i(q_{m}-k_{xn}) L/2} J_{n-m}\left(\frac{V_1}{\hbar\omega}\right)\\
    (M_{cB}^{2})_{nm}=\frac{N'_{m-}}{N_{n+}}e^{-i(q_{m}+k_{xn})L/2} J_{n-m}\left(\frac{V_1}{\hbar\omega}\right)\\
    (M_i)_{nm}=\frac{N_{n+}}{N_{n-}}e^{-i k_{xn} L}\delta_{n,m}\\
    (M_r)_{nm}=(S_{+}(E_n)-S_{-}(E_n))\delta_{n,m}
\end{align}

\section{Scattering matrix}
\label{app2}

In this appendix we show the derivation for the S-
matrix based on the boundary condition.
The scattering matrix is defined as, 
\begin{equation}
    \begin{pmatrix}
        A^o\\
        B^o
    \end{pmatrix}=\begin{pmatrix}
        M_{AA} & M_{AB}\\
        M_{BA} & M_{BB}
    \end{pmatrix}\begin{pmatrix}
        A^i\\
        B^i
    \end{pmatrix}
    =S\begin{pmatrix}
        A^i\\
        B^i
    \end{pmatrix}
\end{equation}
where, $A^i$ and $B^i$ represent the amplitudes of the incident wave and $A^o$ and $B^o$ represent the amplitudes of
the outgoing signal and it can be shown that,
\begin{align}
    M_{AA}=M_{cA}^1 a_A+M_{cA}^2 b_A-M_i\\
    M_{BA}=M_{cB}^1 a_A+M_{cB}^2 b_A
\end{align}   
 with,
\begin{equation}
    \begin{aligned}
        &a_A=[(M_{sb}^+)^{-1}M_{sa}^{+}-(M_{sb}^-)^{-1}-M_{sa}^-]^{-1}[(M_{sb}^+)^{-1}-(M_{sb}^-)^{-1}]M_r\\
    \end{aligned}
\end{equation}
\begin{equation}
    \begin{aligned}
        &b_A=[(M_{sa}^+)^{-1}M_{sb}^{+}-(M_{sa}^-)^{-1}-M_{sb}^-]^{-1}[(M_{sa}^+)^{-1}-(M_{sa}^-)^{-1}]M_r
    \end{aligned}        
\end{equation}

\section{Scattering Matrix}

The scattering matrix is defined as, 
\begin{equation}
    \begin{pmatrix}
        A^o\\
        B^o
    \end{pmatrix}=\begin{pmatrix}
        M_{AA} & M_{AB}\\
        M_{BA} & M_{BB}
    \end{pmatrix}\begin{pmatrix}
        A^i\\
        B^i
    \end{pmatrix}
    =S\begin{pmatrix}
        A^i\\
        B^i
    \end{pmatrix}
\end{equation}

Where, $A^i$ and $B^i$ repesent the amplitudes of the inci-
dent wave and $A^o$ and $B^o$ represent the amplitudes of
the outgoing signal and it can be shown that,
\begin{align}
    M_{AA}=M_{cA}^1 a_A+M_{cA}^2 b_A-M_i\\
    M_{BA}=M_{cB}^1 a_A+M_{cB}^2 b_A\\
\end{align}

With,
\begin{equation}
    \begin{aligned}
        &a_A=[(M_{sb}^+)^{-1}M_{sa}^{+}-(M_{sb}^-)^{-1}-M_{sa}^-]^{-1}[(M_{sb}^+)^{-1}-(M_{sb}^-)^{-1}]M_r\\
    \end{aligned}
\end{equation}
\begin{equation}
    \begin{aligned}
        &b_A=[(M_{sa}^+)^{-1}M_{sb}^{+}-(M_{sa}^-)^{-1}-M_{sb}^-]^{-1}[(M_{sa}^+)^{-1}-(M_{sa}^-)^{-1}]M_r
    \end{aligned}        
\end{equation}

\section{Quasi-bound Energies}
\label{app3}

\textcolor{black}{
In this section, we derive the energy of the quasi-bound state.  These energies correspond to imaginary (real) wave-vectors outside (inside) the potential well resulting in the quasi-bound states to decay exponentially (propagate without decay) outside (inside) the potential well along the $\pm x$-direction. Note that the electrons propagate freely along the $y,z$-directions. To investigate the quasi-bound states, we consider the
wave function as given in Eq. (\ref{eq_wfall}) with $n = m = V_1 = 0$ as follows:
\begin{equation}
	\psi_n=e^{-iEt/\hbar+ik_yy+ik_zz} \\
	\begin{cases}
		\begin{aligned}
			&r N_{-}\begin{pmatrix}1\\ S_-(E)\end{pmatrix}e^{-ik_{x}x} ,&  x<-L/2 \\
			&
			[aN'_{+}\begin{pmatrix} 1\\ S'_+(E) \end{pmatrix}e^{iqx}+b N'_{-}\begin{pmatrix} 1\\ S'_-(E)\end{pmatrix}e^{-iqx}]   ,& \hfill -L/2\leq x\leq L/2\\&
			t N_{+}\begin{pmatrix} 1\\S_+(E)\end{pmatrix}e^{ik_{x}x} ,& x>L/2\end{aligned}\end{cases}
    \end{equation} 
Now, matching the boundary condition at $x =-L / 2$ we get, 
$$
\begin{aligned}
& r N_{-} e^{i k_x L / 2}-aN^{\prime}_+e^{-i q 
L/ 2}-b N_{-}^{\prime} e^{i q L / 2}=0 \\
& r N_-S_{-}(E) e^{i k_x  L / 2}-a N_{+}^{\prime} S_{+}^{\prime}(E) e^{-i q L / 2}-b N^{\prime}_- S_{-}^{\prime}(E) e^{i q L / 2}=0
\end{aligned}
$$
Also, boundary condition at $x=+L/2$ gives - 
$$
\begin{aligned}
& tN_+e^{-i k_{x} L / 2}-a N^{\prime}_+e^{i q L / 2}-b N^{\prime}_-e^{-i q L / 2}=0 \\
& tN_+S_{+}(E) e^{-i k_x L / 2}-a N^{\prime}_+S^{\prime}_+(E) e^{i q L/2}-b N^{\prime}_-S_{-}^{\prime}(E) e^{-i q L/2}=0
\end{aligned}
$$
Secular determinant can be obtained by collecting the coefficients of $r$, $t$, $a$, and $b$
$$
\begin{aligned}
&\left|\begin{array}{cccc}
N_{-} e^{i k_x L / 2} & -N^{\prime}_+e^{-i q 
L/ 2} & -N_{-}^{\prime} e^{i q L / 2} & 0 \\
N_-S_{-}(E) e^{i k_x  L / 2} & -N_{+}^{\prime} S_{+}^{\prime}(E) e^{-i q L / 2} & -N^{\prime}_- S_{-}^{\prime}(E) e^{i q L / 2} & 0 \\
0 & -N^{\prime}_+e^{i q L / 2} & -N^{\prime}_-e^{-i q L / 2} & N_+e^{-i k_{x} L / 2} \\
0 & -N^{\prime}_+S^{\prime}_+(E) e^{i q L/2} & -N^{\prime}_-S_{-}^{\prime}(E) e^{-i q L/2} & N_+S_{+}(E) e^{-i K_x L / 2}
\end{array}\right| =0\\ \\
\end{aligned}
$$
This further yields the following determinant while the basis column matrix is given by $(N_{-} e^{i k_x L / 2}, N^{\prime}_+e^{-i q 
L/ 2}, N_{-}^{\prime} e^{i q L / 2}, N_+e^{-i k_{x} L / 2})^T$ as follows 
$$ \begin{aligned}
& \left|\begin{array}{cccc}
1 & -1 & -1 & 0 \\
S_{-}(E) & -S_{+}^{\prime}(E) & -S_{-}^{\prime}(E) & 0 \\
0 & -e^{i q L} & -e^{-i q L} & 1 \\
0 & -S_{+}^{\prime}(E) e^{i q L} & -S_{-}^{\prime}(E) e^{-i q L} & S_{+}(E)
\end{array}\right| =0
\end{aligned}
$$
Solving the secular determinant and taking only the real part we arrive at - $q_{n}=\frac{(2 n+1) \pi}{2 L}, \hspace{0.5 cm} n \in \mathbb{Z}$\\
From this, we can get the energy of quasi-bound states:
\begin{align*}
    E_{b}= \left[V_{0} \pm \hbar \sqrt{(2 n+1)^{2} \frac{\pi^{2} t^{2}}{4 L^{2}}+t^{2} k_{y}^{2}+s^{2} t_{z}^{2}k_{z}^{2} \sin ^{2} k_{0} }\right]+\hbar k_{z} \tilde{t}
\end{align*}
In our calculations, we consider $\hbar=1$ for simplicity. }

\twocolumngrid



\begin{thebibliography}{0}%
\makeatletter
\providecommand \@ifxundefined [1]{%
 \@ifx{#1\undefined}
}%
\providecommand \@ifnum [1]{%
 \ifnum #1\expandafter \@firstoftwo
 \else \expandafter \@secondoftwo
 \fi
}%
\providecommand \@ifx [1]{%
 \ifx #1\expandafter \@firstoftwo
 \else \expandafter \@secondoftwo
 \fi
}%
\providecommand \natexlab [1]{#1}%
\providecommand \enquote  [1]{``#1''}%
\providecommand \bibnamefont  [1]{#1}%
\providecommand \bibfnamefont [1]{#1}%
\providecommand \citenamefont [1]{#1}%
\providecommand \href@noop [0]{\@secondoftwo}%
\providecommand \href [0]{\begingroup \@sanitize@url \@href}%
\providecommand \@href[1]{\@@startlink{#1}\@@href}%
\providecommand \@@href[1]{\endgroup#1\@@endlink}%
\providecommand \@sanitize@url [0]{\catcode `\\12\catcode `\$12\catcode
  `\&12\catcode `\#12\catcode `\^12\catcode `\_12\catcode `\%12\relax}%
\providecommand \@@startlink[1]{}%
\providecommand \@@endlink[0]{}%
\providecommand \url  [0]{\begingroup\@sanitize@url \@url }%
\providecommand \@url [1]{\endgroup\@href {#1}{\urlprefix }}%
\providecommand \urlprefix  [0]{URL }%
\providecommand \Eprint [0]{\href }%
\providecommand \doibase [0]{http://dx.doi.org/}%
\providecommand \selectlanguage [0]{\@gobble}%
\providecommand \bibinfo  [0]{\@secondoftwo}%
\providecommand \bibfield  [0]{\@secondoftwo}%
\providecommand \translation [1]{[#1]}%
\providecommand \BibitemOpen [0]{}%
\providecommand \bibitemStop [0]{}%
\providecommand \bibitemNoStop [0]{.\EOS\space}%
\providecommand \EOS [0]{\spacefactor3000\relax}%
\providecommand \BibitemShut  [1]{\csname bibitem#1\endcsname}%
\let\auto@bib@innerbib\@empty
\end{thebibliography}%


\begin{thebibliography}{999}

\bibitem{new7} A. A. Burkov, J. Phys. Condens. Matter {\bf 27}, 113201 (2015).

\bibitem{new3} B. Yan and C. Felser, Annu. Rev. Condens. Matter Phys. 8, 337 (2017).

\bibitem{new5} N. P. Armitage, E. J. Mele, and A. Vishwanath, Rev. Mod. Phys. {\bf 90}, 015001 (2018).

\bibitem{new6} E. V. Gorbar, V. A. Miransky, I. A. Shovkovy, P. O. Sukhachov, Low Temperature Physics {\bf 44}, 487 (2018).

\bibitem{arp1} H. Weng, C. Fang, Z. Fang, B. A. Bernevig, and X. Dai, Phys. Rev. X {\bf 5}, 011029 (2015).

\bibitem{arp2} Huang, S., Xu, S., Belopolski, I. et al., Nat. Commun. {\bf 6}, 7373 (2015).

\bibitem{Yan17} B. Yan and C. Felser, Annu. Rev. Condens. Matter Phys. {\bf 8} 337 (2017).

\bibitem{new4} M. Z. Hasan, S.-Y. Xu, I. Belopolski, and C.-M. Huang, Annu. Rev. Condens. Matter Phys. {\bf 8}, 289 (2017).

\bibitem{11} G. Xu, H. Wend, Z. Wang, X. Dai, and Z. Fang, Phys. Rev. Lett. {\bf 107}, 186806 (2011). 

\bibitem{12} C. Fang, M. J. Gilbert, X. Dai, and B. A. Bernevig, Phys. Rev. Lett. {\bf 108}, 266802 (2012).

\bibitem{WSMII} A. A. Soluyanov, D. Gresch, Z. Wang, Q. Wu, M. Troyer, X. Dai, and B. A. Bernevig, Nature {\bf 527}, 495-498 (2015).

\bibitem{FP1} G. Autes, D. Gresch, M. Troyer, A. A. Soluyanov, and O. V. Yazyev, Phys. Rev. Lett. {\bf 117}, 066402 (2016).

\bibitem{FP2} M.-Y. Yao, N. Xu, Q. S. Wu, G. Autes, N. Kumar, V. N. Strocov, N. C. Plumb, M. Radovic, O. V. Yazyev, C. Felser, J. Mesot, and M. Shi, Phys. Rev. Lett. {\bf 122}, 176402 (2019). 



\bibitem{NMR} H. Nielsen and M. Ninomiya, Phys. Lett. B {\bf 130}, 389 (1983).

\bibitem{Nag21}  T. Nag, S. Nandy, J. Phys.: Condens. Matter 33,  {\bf 075504} (2021). 

\bibitem{Sadhukhan23} B. Sadhukhan, T. Nag, Phys. Rev. B {\bf 107}, L081110 (2023).

\bibitem{AM1} A. Menon and B. Basu, J. Phys.: Condens. Matter {\bf 33} 045602 (2021).

\bibitem{Xiong22} F. Xiong, C. Honerkamp, D. M. Kennes, T. Nag, Phys. Rev. B {\bf 106}, 045424 (2022). 

\bibitem{Das21} S. K. Das, T. Nag, S. Nandy, Phys. Rev. B {\bf 104}, 115420 (2021).

\bibitem{Sadhukhan21a} B. Sadhukhan, T. Nag, Phys. Rev. B {\bf 103}, 144308 (2021).

\bibitem{Sadhukhan21} B. Sadhukhan, T. Nag, Phys. Rev. B {\bf 104}, 245122 (2021).

\bibitem{Nag22} T. Nag, D. M. Kennes, Phys. Rev. B {\bf 105}, 214307 (2022). 

\bibitem{T21} T. E. O. Brien, M. Diez, C. W. J. Beenakker, Phys. Rev. Lett. {\bf 116}, 236401 (2016).

\bibitem{T22} Zyuzin, A.A., Tiwari, R.P., JETP Lett. {\bf 103}, 717–722 (2016).

\bibitem{T23} Maximilian Trescher, Björn Sbierski, Piet W. Brouwer, and Emil J. Bergholtz, Phys. Rev. B {\bf 91}, 115135 (2015).

\bibitem{oka09} T. Oka and H. Aoki, Physical Review B {\bf 79}, 081406(R) (2009).

\bibitem{kitagawa10} T. Kitagawa, E. Berg, M. Rudner, and E. Demler, Phys. Rev. B {\bf 82}, 235114 (2010).

\bibitem{rudner13}  M. S. Rudner, N. H. Lindner, E. Berg, and M. Levin, Phys. Rev. X {\bf 3}, 031005 (2013).

\bibitem{nathan15} F. Nathan and M. S. Rudner, New Journal of Physics {\bf 17}, 125014 (2015).

\bibitem{titum16} P. Titum, E. Berg, M. S. Rudner, G. Refael, and N. H. Lindner, Phys. Rev. X {\bf 6}, 021013 (2016).

\bibitem{Nag14} T. Nag, S. Roy, A. Dutta, D. Sen, Phys. Rev. B {\bf 89}, 165425 (2014). 

\bibitem{Nag19}  T. Nag, R-J Slager, T. Higuchi, T. Oka, Phys. Rev. B 100, 134301 (2019).

\bibitem{Tamang21}  L. Tamang, T. Nag, T. Biswas, Phys. Rev. B {\bf 104}, 174308 (2021). 

\bibitem{Nag21b} T. Nag, A. Rajak, Phys. Rev. B {\bf 104}, 134307 (2021).

\bibitem{Kundu21} A. Kundu, A. Rajak, T. Nag,
Phys. Rev. B {\bf 104}, 075161 (2021). 

\bibitem{po16} H. C. Po, L. Fidkowski, T. Morimoto, A. C. Potter, and A. Vishwanath, Phys. Rev. X {\bf 6}, 041070 (2016).

\bibitem{kar18}S. Kar and B. Basu, Phys. Rev. B {\bf 98}, 245119 (2018). 

\bibitem{F1} J. H. Shirley, Phys. Rev. {\bf 138}, B979 (1965).

\bibitem{F2} M. Holthaus and D. Hone, Phys. Rev. B {\bf 47}, 6499 (1993).

\bibitem{F3} T. Fromherz, Phys. Rev. B {\bf 56}, 4772 (1997).

\bibitem{AM2} A. Menon, D. Chowdhury, and B. Basu, Phys. Rev. B {\bf 98}, 205109 (2018).

\bibitem{AM3} T. Nag, A. Menon, and B. Basu, Phys. Rev. B {\bf 102}, 014307 (2020).

\bibitem{FD1} Zhang, B., Maeshima, N., and Hino, Ki., Sci. Rep. {\bf 11}, 2952 (2021).

\bibitem{FQSL} Kumar, U., Banerjee, S., and Lin, S.Z., Nat. Commun. Phys. {\bf 5}, 157 (2022).

\bibitem{Z2} U. Fano, Nuovo Cimento {\bf 12}, 154 (1935).
\bibitem{Z1} A. E. Miroshnichenko, S. Flach, and Y. S. Kivshar, Rev. Mod. Phys. {\bf 82}, 2257 (2010).

\bibitem{Z3} B. Lukyanchuk, N. I. Zheludev, S. A. Maier, N. J. Halas, P. Nordlander, H. Giessen, and C. T. Chong, Nat. Mater. {\bf 9}, 707 (2010).

\bibitem{Z4} E. Tekman and P. F. Bagwell, Phys. Rev. B {\bf 48}, 2553 (1993).

\bibitem{Z5} H. G. Luo, T. Xiang, X. Q. Wang, Z. B. Su, and L. Yu, Phys. Rev. Lett. {\bf 92}, 256602 (2004).

\bibitem{Z6} A. C. Johnson, C. M. Marcus, M. P. Hanson, and A. C. Gossard, Phys.
Rev. Lett. {\bf 93}, 106803 (2004).

\bibitem{Z7} B. R. Bułka and P. Stefanski, Phys. Rev. Lett. {\bf 86}, 5128 (2001).

\bibitem{Z8} M. E. Torio, K. Hallberg, S. Flach, A. E. Miroshnichenko, and M. Titov, Eur. Phys. J. B {\bf 37}, 399 (2004).

\bibitem{Z9} C. S. Chu and R. S. Sorbello, Phys. Rev. B {\bf 40}, 5941 (1989).

\bibitem{Z10} R. Zhu, J. Phys.: Condens. Matter {\bf 25}, 036001 (2013).

\bibitem{Z11} T. B. Boykin, B. Pezeshki, and J. S. Harris, Phys. Rev. B {\bf 46}, 12769 (1992).

\bibitem{Z12} W. Porod, Z. Shao, and C. S. Lent, Appl. Phys. Lett. {\bf 61}, 1350 (1992).

\bibitem{Z13} J. L. Cardoso and P. Pereyra, Europhys. Lett. {\bf 83}, 38001 (2008).

\bibitem{Z14} K. Kobayashi, H. Aikawa, A. Sano, S. Katsumoto, and Y. Iye, Phys. Rev. B {\bf 70}, 035319 (2004).

\bibitem{Z15} R. Zhu and M. Liu, Eur. Phys. J. B {\bf 89}, 2 (2016).

\bibitem{Z16} R. Zhu, J.-H. Dai, and Y. Guo, J. Appl. Phys. {\bf 117}, 164306 (2015).

\bibitem{Z17} R. Zhu, C. Cai, J. Appl. Phys. {\bf 122}, 124302 (2017).

\bibitem{FW1} S. Bera and I. Mandal,  J. Phys.: Condens. Matter {\bf 33}, 295502 (2021).

\bibitem{SB1} S. Bera, S. Sekh, I. Mandal, Ann. Phys. (Berlin) {\bf 535}, 2200460 (2023).


\bibitem{Gregefalk23} A. Gregefalk, A. Black-Schaffer, T. Nag, arXiv:2306.08759 (2023).

\bibitem{Fano1} W. Li and L. E. Reichl, Phys. Rev. B {\bf 60}, 15732 (1999).

\bibitem{Blanter2000} Y. M. Blanter and M. B\"uttiker, Physics reports {\bf 336}, 1 (2000).


\bibitem{Fano61} U. Fano, Phys. Rev. {\bf 124} 1866 (1961).

\bibitem{Milicevic19} M. Milicevic, G. Montambaux, T. Ozawa, O. Jamadi, B. Real, I. Sagnes, A. Lemaitre, L. Le Gratiet, A. Harouri, J. Bloch, and A. Amo, Phys. Rev. X {\bf 9}, 031010 (2019).


\bibitem{40} A. Cortijo, D. Kharzeev, K. Landsteiner, M. A. H. Vozmediano, Phys. Rev. B {\bf 94}, 241405(R) (2016).

\bibitem{55} D. I. Pikulin, Anffany Chen, and M.Franz, Phys. Rev. X {\bf 6}, 041021 (2016).

\bibitem{56} A. G. Grushin, J. W. F. Venderbos, A. Vishwanath, and R. Ilan, Phys. Rev. X {\bf 6}, 041046 (2016).

\bibitem{57} R. W. Bomantara and J. Gong, Phys. Rev. B {\bf 94}, 235447 (2016).

\bibitem{Goerbig08}[M. O. Goerbig, J.-N. Fuchs, G. Montambaux, and F. Piechon, Phys. Rev. B {\bf 78}, 45415 (2008).


\bibitem{Lu16} H-Y Lu, A. S. Cuamba, S-Y Lin, L. Hao, R. Wang, H. Li, Y. Zhao, and C. S. Ting, Phys. Rev. B {\bf 94}, 195423 (2016).

\bibitem{Zabolotskiy16}A. D. Zabolotskiy and Yu. E. Lozovik, Phys. Rev. B {\bf 94}, 165403 (2016).


\bibitem{Polozkov19} R. G. Polozkov, N. Y. Senkevich, S. Morina, P. Kuzhir, M. E. Portnoi, and I. A. Shelykh, Phys. Rev. B {\bf 100}, 235401 (2019).

\bibitem{Lang23} J-P. Lang, H. Hanafi, J. Imbrock, and C. Denz, Phys. Rev. A {\bf 107}, 023509 (2023).



\bibitem{Shore67} B. W. Shore, Rev. Mod. Phys. {\bf 39}, 439 (1967).

\bibitem{Smith60} F. Smith, Phys. Rev. {\bf 118}, 349 (1960).

\end{thebibliography}
\end{document}